\documentclass[12pt,amssymb]{article} 

\usepackage{mathrsfs}
\usepackage{amssymb}
\usepackage[dvips]{graphicx}

\newcommand{\newsection}[1]{
\addtocounter{section}{1} \setcounter{equation}{0}
\setcounter{subsection}{0} \addcontentsline{toc}{section}{\protect
\numberline{\arabic{section}}{{\rm #1}}} \vglue .6cm \pagebreak[3]
\noindent{ \bf  \thesection. #1}\nopagebreak[4]\par\vskip .3cm}
\newcommand{\newsubsection}[1]{
\addtocounter{subsection}{1}\setcounter{subsubsection}{0}
\addcontentsline{toc}{subsection}{\protect
\numberline{\arabic{section}.\arabic{subsection}}{#1}} \vglue .4cm
\pagebreak[3] \noindent{\it \thesubsection.
#1}\nopagebreak[4]\par\vskip .3cm}

\makeatletter
\newcommand{\seclabel}[1]{%
  \@bsphack
  \protected@write\@auxout{}%
     {\string\newlabel{#1}{{\thesection}{\thepage}}}
  \@esphack
  }
\newcommand{\subseclabel}[1]{%
  \@bsphack
  \protected@write\@auxout{}%
     {\string\newlabel{#1}{{\thesubsection}{\thepage}}}
  \@esphack
  }
\newcommand{\tablabel}[1]{%
  \@bsphack
  \protected@write\@auxout{}%
     {\string\newlabel{#1}{{\arabic{tabnum}}{\thepage}}}
  \@esphack
  }
\makeatother

\renewcommand{\theequation}{\thesection.\arabic{equation}}

\newlength{\extraspace}
\newlength{\extraspaces}
\newcounter{dummy}
\newcommand{\bc}{\begin{center}}
\newcommand{\ec}{\end{center}}
\newcommand{\be}{\begin{equation}
\addtolength{\abovedisplayskip}{\extraspaces}
\addtolength{\belowdisplayskip}{\extraspaces}
\addtolength{\abovedisplayshortskip}{\extraspace}
\addtolength{\belowdisplayshortskip}{\extraspace}}
\newcommand{\ee}{\end{equation}}

\newcommand{\ba}{\begin{eqnarray}
\addtolength{\abovedisplayskip}{\extraspaces}
\addtolength{\belowdisplayskip}{\extraspaces}
\addtolength{\abovedisplayshortskip}{\extraspace}
\addtolength{\belowdisplayshortskip}{\extraspace}}
\newcommand{\ea}{\end{eqnarray}}

\newcommand{\ban}{\begin{eqnarray*}
\addtolength{\abovedisplayskip}{\extraspaces}
\addtolength{\belowdisplayskip}{\extraspaces}
\addtolength{\abovedisplayshortskip}{\extraspace}
\addtolength{\belowdisplayshortskip}{\extraspace}}
\newcommand{\ean}{\end{eqnarray*}}
\newcommand{\baa}{
\addtocounter{equation}{1} \setcounter{dummy}{\value{equation}}
\setcounter{equation}{0}
\renewcommand{\theequation}{\thesection.\arabic{dummy}\alph{equation}}
\begin{eqnarray}
\addtolength{\abovedisplayskip}{\extraspaces}
\addtolength{\belowdisplayskip}{\extraspaces}
\addtolength{\abovedisplayshortskip}{\extraspace}
\addtolength{\belowdisplayshortskip}{\extraspace}}
\newcommand{\eaa}{
\end{eqnarray}
\setcounter{equation}{\value{dummy}}
\renewcommand{\theequation}{\thesection.\arabic{equation}}}

\input epsf

\newcounter{fignum}
\newcounter{tabel}

\newcounter{tabnum}
\newcommand{\vev}[1]{\left\langle #1\right\rangle}

\newcommand{\half}{\frac{1}{2}}
\newcommand{\del}{\partial}
\newcommand{\delb}{\bar{\del}}
\newcommand{\eol}{\nonumber \\}

\newcommand{\cO}{{\cal O}}
\newcommand{\Ext}{{\rm Ext}}
\newcommand{\Hom}{{\rm Hom}}

\begin{document}

%
%

\begin{flushright}
April 2011\\
%
\end{flushright}
\vspace{1.5cm}

\thispagestyle{empty}

%
%

\begin{center}
{\Large\bf  Gluing Branes, I
 \\[13mm] }

{\sc Ron Donagi}\\[2.5mm]
{\it Department of Mathematics, University of Pennsylvania \\
Philadelphia, PA 19104-6395, USA}\\[9mm]

{\sc Martijn Wijnholt}\\[2.5mm]
{\it Arnold Sommerfeld Center, Ludwig-Maximilians Universit\"at\\
Theresienstrasse 37 \\
D-80333 M\"unchen, Germany }\\
[25mm]

 {\sc Abstract}

\end{center}

We consider several aspects of holomorphic brane configurations. We
recently showed that an important part of the defining data of such
a configuration is the gluing morphism, which specifies how the
constituents of a configuration are glued together, but is usually
assumed to be vanishing. Here we explain the rules for computing
spectra and interactions for configurations with non-vanishing
gluing VEVs. We further give a detailed discussion of the $D$-terms
for Higgs bundles, spectral covers and ALE fibrations. We highlight
a stability criterion that applies to degenerate configurations of
the spectral data, and address an apparent discrepancy between the
field theory and ALE descriptions. This allows us to show that one
gets walls of marginal stability in $F$-theory even though they are
absent in the $11d$ supergravity description. We also propose a
numerical approach for approximating the hermitian-Einstein metric
of the Higgs bundle using balanced metrics.

\vfill

\newpage

\renewcommand{\Large}{\normalsize}

\tableofcontents

\newpage

\newsection{Introduction}

\newsubsection{Gluing branes}

Brane configurations play a central role in string theory. The low
energy worldvolume theory of smooth weakly curved branes is usually
described by a dimensionally reduced version of the $10d$
supersymmetric Yang-Mills theory. In order to engineer a wider class
of Yang-Mills theories, we can consider configurations which are not
quite smooth. The prime example is to consider intersecting branes,
in order to get charged matter.

The main purpose of this paper is to revisit some very basic
properties of such degenerate brane configurations. They will be
mostly holomorphic, although we will also make some comments on
branes that are not of this type.

Consider first a pair of intersecting $D$-branes. At first sight one
might think that such a configuration is specified by writing down
holomorphic cycles $D_1,D_2$ and holomorphic line bundles $L_1,L_2$
on each of them. However in \cite{Donagi:2010pd} we showed that this
data is incomplete. In addition, one has to specify how the line
bundles are glued along the intersection $D_1 \cap D_2$. This gluing
data is given by a birational isomorphism between the line bundles
along the intersection, i.e. a meromorphic map between
$L_1|_{D_1\cap D_2}$ and $L_2|_{D_1\cap D_2}$. It is usually
implicitly assumed that this gluing morphism vanishes, but this is
non-generic.

The gluing morphism also gives a new perspective on brane
recombination. When expanding around an intersecting configuration
with vanishing gluing morphism, given by $xy=0$ say, one finds
massless modes $Q$ and $\tilde Q$ with opposite $U(1)$-charges at
the intersection. A non-zero VEV for $\left< \right.\!Q\tilde
Q\!\left. \right>$ leads to a smoothing of the brane intersection,
of the form $xy \sim \left< \right.\!Q\tilde Q\!\left. \right>$.
However when embedded in more complicated configurations, such a VEV
is often disallowed by the $F$-term equations, and one is interested
in deformations with $\vev{Q}\not = 0$ and $\left<\right.\!\tilde
Q\!\left.\right>=0$. So the question arises how to interpret this
geometrically. We found that at the level of $F$-terms, turning on
$\vev{Q}$ can be represented by turning on a gluing VEV, without
changing the support of the branes \cite{Donagi:2010pd}.

These observations raise a number of new questions about
intersecting brane configurations. For most of this paper, we will
be interested in configurations where the gluing morphism does not
vanish. We will explain how to compute the spectrum and interactions
in such cases, and we will discuss aspects of the $D$-terms. We will
see how the above observations resolve several puzzles about
intersecting branes. For example, the low energy theory around a
point of $U(1)$ restoration is believed to be described by the Fayet
model. But if the brane intersection were somehow smoothed out by
the VEV for $\vev{Q}$, then this could not be correct, because line
bundles on smooth divisors are always stable. In addition, it would
not be compatible with $T$-duality/Fourier-Mukai transform. Our
results naturally resolve these problems.

It is frequently useful to regard intersecting configurations as a
limit of smooth configurations, which are more generic. There are
many other interesting types of degenerations. Apart from
intersecting branes, one of the simplest possibilities is a
holomorphic cycle that has some multiplicity. Such configurations
are said to be non-reduced. It is usually assumed that a rank one
sheaf over a non-reduced cycle $r D$ takes the form of a rank $r$
vector bundle over $D$. However it is known that there are other
possibilities, namely sheaves that are non-trivial on the
infinitesimal neighbourhoods of $D$. These were first studied
mathematically in \cite{DEL} in the context of Higgs bundles and
their deformations. Sheaves on non reduced schemes appeared in a
string-theoretic context in \cite{Aspinwall:1998he}, where their
moduli space was analyzed for a compactification on $K3$. The first
explicit, systematic appearance in physics of non-diagonalizable
Higgs fields and the related sheaves on non-reduced schemes was in
\cite{Donagi:2003hh}. The local structure of such non-reduced
schemes is identical to the above structure over brane
intersections. Such configurations have recently also been studied
in \cite{Cecotti:2010bp,Chiou:2011js}. The possibility that the
sheaf takes the form of a non-trivial rank $r$ bundle over $D$ is
of course also interesting, and has been studied in for example
\cite{Bershadsky:1997ec,Bershadsky:1997zs,Berglund:1998ej,Aspinwall:2000kf}.
Although such configurations are not the focus of the present
paper, they are easily included as special cases and in this paper
and its follow-up we will see explicitly how to calculate with
general configurations that include all the ingredients above.

The degenerations we study in this paper are in some sense the
simplest ones, and they do not exhaust the list of possibilities. It
would be of interest to get some kind of classification of the
allowed degenerations. We also emphasize that our discussion applies
to holomorphic branes generally, whether they appear in the context
of $F$-theory, the heterotic string or perturbative type IIb. In
fact, much of the story also appears to work for $A$-branes. This
looks particularly promising for $M$-theory phenomenology, as one
may try to construct models with bulk matter and classical Yukawa
couplings. Until now, Yukawa couplings in such models were induced
by instanton effects, and thus rather small.

\newsubsection{D-terms}

We would also like to take the opportunity to address some questions
involving the $D$-terms in $F$-theory. One issue which has bothered
us for some time is an apparent discrepancy between the $D$-term
equations in the worldvolume and the space-time descriptions. In the
space-time description, Becker and Becker \cite{Becker:1996gj} found
that for smooth Calabi-Yau four-folds the $D$-terms are given by a
primitiveness condition, viz. $J \wedge G = 0$. Although the
geometries of interest for engineering gauge theories are not
smooth, one might have thought that some version of this equation
holds for singular Calabi-Yau four-folds, by first resolving and
then taking a limit.

However there are several problems with this idea. In interesting
cases, the resolution of the four-fold can be obstructed by the
background three-form field. Furthermore in the brane worldvolume
description, solutions to the $D$-term equations correspond to
Higgs bundles that are stable. This condition is manifestly not
equivalent to $J \wedge G = 0$, because primitiveness is a closed
condition and stability is an open condition. (More precisely, the
correct condition is poly-stability, which is locally closed and
therefore still inequivalent). And as a related problem, the
Fayet-Iliopoulos parameters in $F$-theory are given by expressions
of the form $\int G \wedge J \wedge \omega$, which we would expect
could be non-zero in regimes where the supergravity approach of
\cite{Becker:1996gj} does not apply. But then we clearly should not
impose $J \wedge G = 0$. So what is then the correct version of the
$D$-term equation on a Calabi-Yau four-fold?

As discussed in section \ref{hEinsteinMetric}, such a situation was already
encountered in \cite{Wijnholt:2001us}, and it works
exactly the same way here. Namely the condition $J \wedge G=0$ must
be corrected for singular or close-to-singular Calabi-Yau
four-folds, when non-abelian degrees of freedom are light, but the
non-abelian corrections cannot be properly incorporated in this
picture. To study physical wave-functions and other properties of
the $D$-terms, we must use the Higgs bundle picture, as it is the
only picture in which the non-abelian degrees of freedom are
properly included. We note that this yields another rationale for
the strategy of splitting the study of $F$-theory (or $M$-theory, or
type I') into local and global models.

Despite this, we further argue that there is still a sense in which
we can include the non-abelian corrections even in the Calabi-Yau picture,
by replacing the primitiveness condition of
\cite{Becker:1996gj} by a notion of slope stability for four-folds
with flux. Stability makes sense at the level of algebraic geometry
and should preserve the essential information of existence and
uniqueness of a solution in the Higgs bundle picture. At any rate, in the
regime of $F$-theory where the $8d$ gauge theory is weakly coupled, we find
a chamber structure on the K\"ahler moduli space with walls of marginal stability,
exactly as expected in the general context of geometric
invariant theory \cite{ThadGITVar,DolgHu} and observed in heterotic
models. (Such a chamber structure was also expected for intersecting
branes in type II, but as noted above, our picture of brane
recombination is needed to realize it). Such a structure would be
difficult to explain with a primitiveness condition.

Another issue that we would like to address is the actual
computation of physical wave functions and terms in the K\"ahler
potential. It has been hard to get a handle on this due to the
difficulty of solving the $D$-term equations explicitly. But it is
also crucial for getting a more detailed understanding of the
$D$-terms for degenerate cases and for issues such as dimension six
proton decay. In section \ref{NumMetric} we will explain a possible
procedure for numerically approximating the solutions of the
$D$-term equations of Higgs bundles using balanced metrics.

Finally, in section \ref{SimpsonStability} we discuss how to
formulate the criterion for existence of solutions to the $D$-terms
directly in terms of the spectral data. We highlight the notion of
stability for sheaves which applies even to configurations where the
spectral cover is degenerate. This connects the discussion of the
$D$-terms with the rest of the paper.

The present paper is the first of two papers on degenerate brane
configurations, and focusses on theoretical aspects. Part II
contains applications to heterotic/$F$-theory duality for gauged
linear sigma models and to model building. There we discuss how to
engineer models with matter in the bulk of a brane and with various
flavour structures, without generating exotics. In particular, we
address the issue of proton decay, and describe a solution to the
mu-problem which puts the Higgs fields in the bulk and does not use
a $U(1)$ gauge symmetry.

\newpage

\newsection{Degenerate Branes}

\seclabel{DegenerateBranes}

\newsubsection{Higgs bundles versus spectral covers}

\subseclabel{HiggsvsSpectral}

Before we discuss degenerate configurations, it will be helpful to
recall some general aspects of Higgs bundles and their relation to
$8d$ supersymmetric Yang-Mills theory
\cite{DonagiMarkmanReview,Hayashi:2009ge,Donagi:2009ra}. Pieces of
this story were also worked out in
\cite{Donagi:2008ca,Beasley:2008dc}.

The worldvolume theory of a brane is the maximally supersymmetric
Yang-Mills theory with gauge group $G$. For concreteness we consider
the eight-dimensional Yang-Mills theory, though analogous statements
can be made in other dimensions. The bosonic fields are given by a
gauge field $A$ on a bundle $E$, and a complex adjoint field $\Phi$.
The Yang-Mills Lagrangian is unique, but when the brane is curved
the higher derivative corrections may become important. We will
always assume that the brane is weakly curved so that we can ignore
the higher order corrections, which we typically wouldn't know how
to calculate anyway.

When the gauge theory is compactified on a complex surface $S$, and
we insist on preserving $N=1$ supersymmetry in $4d$, then the
adjoint field is twisted by the canonical bundle of $S$. So the
bosonic fields take values in
\be A\ \in\ \Omega^1({\rm ad}({\cal G})), \qquad \Phi^{2,0}\ \in\
\Omega^0({\rm Ad}({\cal G}) \otimes K_S) \ee
where ${\cal G}$ is the bundle of frames associated to $E$. We will
often denote $\Phi^{2,0}$ simply by $\Phi$. These fields have to
satisfy the $F$-term equations:
\be F^{0,2}\ =\ 0, \qquad \delb_A\Phi^{2,0}\ =\ 0 \ee
As is well-known, $F^{0,2}=0$ implies that the bundle is
holomorphic, and we can then further simplify by choosing a
non-unitary gauge such that $\delb_A =\delb$. Solutions of these
equations define a $K_S$-twisted Higgs bundle. The $D$-terms are
discussed in section \ref{Dterms}. In the following, we will take
the gauge group to be $U(r)$.

It is convenient to reinterpret $\Phi$ in the following way. Let us
denote the total space of the canonical bundle by $X$, and the
projection $X \to S$ by $\pi$. Given such data, a standard
construction known as the Higgs/spectral correspondence rewrites the
holomorphic data as a spectral sheaf on $X$. Let $\lambda$ be the
canonical section of $\pi^*K_S$ which vanishes on the zero section.
We may identify the conormal bundle $K_S^{-1}$ with $I/I^2$, where
$I = \vev{\lambda}$ is the ideal generated by $\lambda$. Also let
$m$ a local section of $E$. Then we may define an action of
$\lambda$ on $E$ in the following way:
\be\label{HiggsAction} \lambda \cdot m \ = \ \Phi m \ee
Since $\Phi \wedge \Phi = 0$, it follows that $E$ can be regarded as
a module over the symmetric algebra ${\rm
Sym}^\bullet(\cO_S(K_S^{-1})) = \cO_X$, and hence defines a sheaf
${\cal L}$ on $X$. So as far as the $F$-terms are concerned, a Higgs
bundle on $S$ is tautologically the same as a coherent sheaf on $X$,
whose support is of pure dimension ${\rm dim}(S)$, and finite over
$S$.

More geometrically, let us interpret $\Phi$ as a holomorphic map
\be \Phi:\ E\ \to\ E \otimes K_S \ee
Denote by $\pi$ the projection $X \to S$, and let us consider the
bundles $\pi^*E$ and $\pi^*E \otimes K_S$ on $X$. We have a map
\be \Psi \equiv \lambda I - \Phi: \ \pi^*E\ \to \ \pi^*(E \otimes
K_S) \ee
where $\lambda$ is the canonical eigenvalue section as above. As a
map between sheaves this is injective, because on open subsets of
$X$ it has rank $r$. Then (\ref{HiggsAction}) is equivalent to
saying that we define the spectral sheaf ${\cal L}$ as the cokernel
of the map $\Psi$. In other words, the spectral sheaf is defined
through an exact sequence
\be\label{CokerDef} 0 \ \to  \ \pi^*E\ \to \ \pi^*(E \otimes K_S) \
\to \ {\cal L} \ \ \to\ 0 \ee
To get some intuition, let us view this construction more locally.
Generically the eigenvalues are distinct, and thus we may use a
complexified gauge transformation to diagonalize $\Phi$. Then we get
\be\label{HBundleMap} \lambda I - \Phi \ \sim \left(
  \begin{array}{ccc}
    \lambda -\lambda_1 & 0 & 0 \\
    0 & \ddots & 0 \\
    0 & 0 & \lambda-\lambda_r \\
  \end{array}
\right) \ee
For generic $\lambda$ the map $\lambda I - \Phi$ has rank $r$, and
thus the cokernel vanishes. However on a sublocus the rank drops to
$r-1$, and the cokernel is one-dimensional. Thus ${\cal L}$
generically looks like a line bundle supported on the spectral
cover, which is the holomorphic divisor $C$ in $X$ defined by the
equation
\be \det(\lambda I - \Phi) = 0 \ee
More precisely, ${\cal L}$ is a rank one sheaf, where by rank we
mean the coefficient of the leading term in the Hilbert polynomial.
The spectral sheaf and the Higgs bundle are equivalent, at least as
far as the holomorphic data is concerned. We saw above how to
construct a spectral sheaf out of the Higgs bundle. Conversely,
given a spectral sheaf, we can construct a Higgs bundle as $E
\otimes K \simeq p_{C*}{\cal L}$ and $ \Phi \simeq p_{C*}\lambda I$,
where $p_C$ is the covering map $p_C: C \to S$.

In \cite{Donagi:2009ra,Pantev:2009de,PW-p} we further argued that
such constructions are equivalent to supersymmetric ALE-fibrations,
through a version of the cylinder mapping. In this form, they can be
pasted into compact models in $F$-theory. The same strategy can also
be employed in $M$-theory and type I'.

For $SU(n)$ bundles, the support of the spectral sheaf is
generically a smooth complex surface. This follows from Bertini's
theorem, which says that the generic element of a linear system is
smooth and irreducible. In this paper we will be interested in some
of the simplest degenerations of such smooth configurations. Namely
we will consider degenerations where the divisor becomes reducible
or non-reduced. It should be emphasized that the correspondence
reviewed above is tautological. It is irrelevant whether we consider
a smooth spectral surface or the degenerate cases in this paper. To
trust the Yang-Mills theory physically, we need $\Phi$ and its
derivatives to remain small. This also depends on the hermitian
metric solving the $D$-terms.

Since we have two equivalent ways to represent the same
(holomorphic) data, there will be two equivalent ways to calculate
the spectrum and the holomorphic couplings
\cite{DonagiMarkmanReview,Donagi:2009ra}. On the one hand, we can
use a Dolbeault operator modified by the Higgs VEV:
\be \bar D \ = \ \delb_A + \Phi^{2,0} \ee
Let us define the two-term complex
\be {\cal E}^{\, \bullet} \ = \ {\rm ad}(E)\ \mathop{\to}^{{\rm
ad}(\Phi)}\ {\rm ad}(E) \otimes K \ee
Then to find the massless modes we are interested in the cohomology
of $\bar D$ acting on the spinor configuration space,
$\bigoplus_{p,q}\Omega^{0,p}(S, {\rm ad}(E) \otimes \Lambda^q K_S)$.
This is precisely the hypercohomology $\mathbb{H}^{p+q}({\cal E}^{\,
\bullet})$ of the complex ${\cal E}^{\, \bullet}$. In general, the
unbroken symmetry generators are computed by $\mathbb{H}^0({\cal
E}^{\, \bullet})$, and the massless chiral fields are counted by
$\mathbb{H}^1({\cal E}^{\, \bullet})$. Similarly, the Yukawa
couplings are computed by the Yoneda product on $\mathbb{H}^1({\cal
E}^{\, \bullet})$, and higher order holomorphic couplings by the
higher Massey products on $\mathbb{H}^1({\cal E}^{\, \bullet})$.

On the other hand, we can represent the Higgs bundle configuration
by spectral data and use standard algebraic machinery to compute the
unbroken symmetries, the infinitesimal deformations and their
interactions, which are computed by $\Ext$ groups according to the
deformation theory of sheaves. (See for example
\cite{HartshorneDef}). These two points of view are equivalent.
After expressing the Higgs bundle data by spectral data and using a
spectral sequence argument, we get
\be\label{SpecIso} \mathbb{H}^p({\cal E}^{\, \bullet}) \ = \
\Ext^p_X({\cal L},{\cal L}) \ee
The latter perhaps obscures some geometric intuition, particularly
regarding the $D$-terms, but is more powerful in actual
calculations, because the spectral data is an `abelianized'
presentation of the non-abelian Higgs bundle. Again we emphasize
that this applies quite generally, even when the spectral data is
not a line bundle but merely a coherent sheaf, and independent of
whether the hermitian metric solves the $D$-terms or not.

With this formulation we can further give a concise description of
some of the results in \cite{Donagi:2008ca} (see also
\cite{Hayashi:2009ge,Donagi:2009ra}). In \cite{Donagi:2008ca} the
right-hand side of (\ref{SpecIso}) (and its associated holomorphic
couplings) arose on the heterotic side after Fourier-Mukai
transform, and the left-hand side of (\ref{SpecIso}) (and its
associated holomorphic couplings) arose in the proposed $8d$ field
theory description on the $F$-theory side. So the claim of
\cite{Donagi:2008ca} was that these two expressions naturally
agree.

\newsubsection{Parabolic Higgs bundles and surface operators}
\subseclabel{ParHiggs}

We will typically be interested in Higgs bundles on manifolds like
${\bf P}^1$ or a del Pezzo surface, whose canonical bundle is
negative. Then the canonical bundle and its powers do not have
sections, and so the reader may wonder whether the set of possible
Higgs field configurations is going to be very limited. This
concern goes away if we recall that the Higgs bundles involved may
be meromorphic, i.e. they are valued not in the canonical bundle
$K$ but in a twisted version $K(D)$ for some divisor $D$. For
suitable $D$ the line bundle $K(D)$ and its powers can easily have
sections. Intuitively in a brane picture, this corresponds to
allowing for non-compact `flavour branes.'

Thus in applications one often needs to worry about boundary
conditions. The specific meromorphic Higgs bundles that occur in
$F$-theory have apparently not been much discussed in the
mathematical literature, but closely related structures have been
studied in much detail. A common type of meromorphic Higgs bundle
is a quasi-parabolic Higgs bundle, see eg. \cite{SimpsonHarmonic}.
Below we will review what this entails, in order to illustrate the
type of structure that one will also encounter for $F$-theoretic
Higgs bundles. We will try to be relatively brief, because we will
not explicitly use them in the present paper. In recent years, the
study of such defects has also been picked up in the physics
literature under the name of surface operators.

Let $D \subset S$ be an effective divisor. A quasi-parabolic bundle
is a bundle $E$ on $S$ and a choice of filtration at $D$:
\be  E|_D = E_{0}\ \supset\ \ldots\ \supset\ E_{n} =0 \ee
More generally, one requires a reduction of the structure group to
a parabolic subgroup along $D$. If we also have a meromorphic Higgs
field compatible with the filtration, i.e. such that
\be \Phi(E_i)\ \subset\ E_{i}\otimes K(D) \ee
then it is a quasi-parabolic Higgs bundle. It is called tame if the
Higgs field has at most simple poles, and wild if there are poles of
higher order. It does not appear necessary to impose tameness in
$F$-theory, but we assume this in the following.

The above boundary data should be viewed as complex structure
moduli. There is additional boundary data one should specify, as the
$(1,1)$ part of the curvature may have singularities along $D$:
\be F^{1,1}\ \sim \ 2\pi \alpha\,\delta^2_D + \ldots \ee
A priori one might think that $\alpha$ should be $z$-dependent,
where $z$ is a coordinate along $D$, as the connection is not flat.
However it appears that solutions actually have $\alpha$ constant
along $D$ \cite{SibnerSibner,Rade}. In this case, a choice of
$\alpha$ is the same as a choice of weights for each step in the
filtration above, and a quasi-parabolic bundle with a choice of
weights is called a parabolic bundle. The degree of the bundle
\be {\rm deg}(E)\ =\ \int_S J \wedge {i\over 2\pi} {\rm Tr}(F^{1,1})
\ee
effectively gets a contribution localized along $D$, and so the
slope depends on the boundary data. The Higgs field itself does not
contribute to the degree as ${\rm Tr}([\Phi^\dagger,\Phi])=0$.
Although this extra data does not affect the holomorphic structure,
it plays a role in the $D$-terms through the stability condition.
Effectively it yields additional K\"ahler moduli, and the
hermitian-Einstein metric will depend on these K\"ahler moduli.
In the context of conventional parabolic Higgs bundles on curves, it is
known that varying the weights can induce birational transformations
on the moduli space. Furthermore, the weights should be complexified
by adding a theta-angle $\eta$ \cite{Gukov:2006jk}, which introduces
an extra phase for each configuration in the path integral.

The definitions can be naturally generalized to principal Higgs
bundles with any gauge group $G$. The choice of $\alpha$ determines
a parabolic subgroup $P \subset G_{\mathbb C}$, which takes the
place of the filtration above. The subset $L = P\cap G$ is the
subgroup of $G$ that commutes with $\alpha$ and hence is left
unbroken by the surface operator. The broken gauge generators lead
to an effective gauged sigma model along $D$ with target given by
the coset $G/L = G_{\mathbb C}/P$. This is extended to
$T^*(G_{\mathbb C}/P)$ if we include broken symmetries of the Higgs
field. When $G_{\mathbb C}/P$ admits a linear sigma model
construction, we can think of this as introducing some charged
hypermultiplet degrees of freedom on $D$ which are not part of the
$8d$ gauge theory and turning on Fayet-Iliopoulos terms
(corresponding to $\alpha$), thus giving a VEV to the hypers. The
non-linear description however is more general.

The spectral correspondence extends to quasi-parabolic Higgs
bundles. We compactify $X$ to
\be \bar X\ =\ {\bf P}(\cO_S \oplus K_S) \ee
and we compactify $C$ to $\bar C$ by adding the divisor at
infinity. In the mathematics literature, the spectral cover is
instead often embedded in
\be X' \ = \ {\bf P}(\cO_S \oplus K_S(D)) \ee
where it does not intersect infinity. These two constructions are
related by a birational transformation, so they contain the same
information at the level of $F$-terms. The birational transformation
consists of blowing up $X'$ along $D$ and blowing down the ${\bf
P}^1$ fibers of $X'$.

Denote by $\cO_{X'}(1)$ the line bundle which restricts to $\cO(1)$
on the ${\bf P}^1$-fibers. Introducing homogeneous coordinates
$(s_0,s_1)$ on the fiber of $X'$, we extend the map $\Psi$ to
\be \Psi  =  s_1 I - s_0 \Phi: \ \pi^*E\ \to\ \pi^*E \otimes K(D)
\otimes \cO_{X'}(1) \ee
and define the spectral sheaf as the cokernel. It is localized on
$\bar C = \{\det(\Psi)=0\}$. A quasi-parabolic structure on the
Higgs bundle yields a quasi-parabolic structure on the spectral
sheaf. Namely, we get a filtration by coherent subsheaves
\be  {\cal L}=\mathcal{F}_{0}{\cal L}\ \supset\ \ldots\ \supset\
\mathcal{F}_n{\cal L}  \ee
where $\mathcal{F}_{n}{\cal L}={\cal L} \otimes \cO_X(-\pi^*D)$.
Conversely, given a filtered spectral sheaf ${\cal L}$ we get a
Higgs bundle by pushing down as before.

For physical applications we need to understand the deformation
theory of such Higgs bundles. We want to determine the endomorphisms
and deformations which are normalizable with respect to the
$L^2$-norm defined by the hermitian-Einstein metric. See section
\ref{Dterms} for more information on this. We should be able to give
an algebraic characterization of such modes. Markman
\cite{MarkmanInf} (see also \cite{DonagiMarkmanReview}) and Yokogawa
\cite{YokogawaInfinitesimal} have defined hypercohomology groups for
quasi-parabolic Higgs bundles. These would seem to be natural
candidates for computing the normalizable modes, but this does not
seem to have been worked out. For work in this direction, in the
case of cotangent twisted Higgs bundles, see Mochizuki
\cite{Mochizuki}. Yokogawa also generalizes $\Ext$-groups to
parabolic Higgs sheaves. These should be isomorphic to the
hypercohomology groups of the Higgs bundle under the Higgs
bundle/spectral cover correspondence.

In practice, we are mostly interested in charged chiral matter. This
appears to be well-localized, and so we can be somewhat cavalier
about the precise cohomology groups that one needs.

It is interesting to note that mathematicians have used parabolic
Higgs bundles with rational weights to describe Higgs bundles on
orbifold spaces. According to \cite{Donagi:2008ca}, $F$-theory duals
of heterotic models with discrete Wilson lines (and no exotic
matter) have orbifold singularities, at least in the stable
degeneration limit. Thus it might be interesting to understand if
such parabolic Higgs bundles could be used to describe duals of discrete
Wilson lines, i.e. if this is the correct surface operator to consider
from the point of view of heterotic/$F$-theory duality.
A number of issues would need to be clarified. One might also speculate that
we could generate this surface operator by integrating out heavy
charged states on the heterotic side.

\newsubsection{Structure sheaf of a fat point}

\subseclabel{FatPoints}

 We would like to take the opportunity to
introduce the structure sheaf of a fat point, and analyze it from
several different points of view. This will be the model for the
degenerate cases we consider, so we will encounter the same basic
structure many times over.

It may be helpful to briefly review some of the basics of scheme
theory. The discussion will be local, i.e. we consider $X = {\bf
C}^3$. Essentially all that we will need is described in the next
two paragraphs.

Roughly speaking, a scheme is an algebraic variety, except that we
can have nilpotent elements in the coordinate ring, whereas for an
algebraic variety there are no nilpotents. The simplest example is
to take the complex line ${\bf C}[\epsilon]$, and consider the
equation $\epsilon^2=0$. This defines a double point, or fat point
of length two. Its coordinate ring contains an infinitesimal
generator $\epsilon$ such that $\epsilon^2 = 0$. If the coordinate
ring has such nilpotent elements, then the scheme is said to be
non-reduced. Given a non-reduced scheme $R$, there is an associated
reduced scheme $R_{red}$, and a natural restriction map
\be \cO_{R_{}} \ \to \ \cO_{R_{red}} \ee
obtained by setting all the nilpotent elements to zero.

On any open set $U$, we may consider the collection of local
holomorphic functions over $U$. They fit together in a global object
which is called the structure sheaf $\cO$. We are interested in
sheaves of modules over $\cO$. That is, over any open set $U$, it is
a module $M_U$ over the set of local holomorphic functions $f_U$,
\be f_U \cdot M_U \subset M_U \ee
We will be interested in well-behaved sheaves, which should satisfy
some extra properties. For instance, we will want $M_U$ to be
finitely generated.

A nice way to see non-reduced structures arise is by considering the
fibre-wise behaviour of a Higgs bundle at the ramification locus
(again see \cite{DonagiMarkmanReview} for review).
 Let us consider a simple spectral cover with
equation $\lambda^2 - z = 0$, where as usual $z$ is a coordinate on
the base and $\lambda$ is a coordinate on the fiber of $K_S$. At
$z=0$ this reduces to the equation of a fat point, $\lambda^2 = 0$.

Now we take the trivial line bundle $\cO$ on $z-\lambda^2 = 0$, and
consider the Higgs bundle $E = p_{C*}\cO$. Away from the branch
locus, this is clearly isomorphic to $\cO \oplus \cO$, with a
diagonal Higgs field
\be \Phi \ = \
\left(
  \begin{array}{cc}
    \sqrt{z} & 0 \\
    0 & -\sqrt{z} \\
  \end{array}
\right)  \ee
At $z=0$ it looks like two coinciding branes, so a priori one
possibility is that $E$ is isomorphic to $\cO \oplus \cO$ with
diagonal Higgs field even there. However this is not compatible with
$\delb \Phi = 0$. Moreover the structure sheaf of $z - \lambda^2=0$
is simply the sheaf of sections of a smooth line bundle, so the
space of eigenvectors of $\Phi$ must be one-dimensional at $z=0$,
not two-dimensional.

Let us consider the structure near the ramification locus in more
detail. We have
\be p_{C*}\cO\ =\ \cO_+ \oplus \cO_- \ee
where $\cO_+$ consists of functions which are even under $\lambda
\to -\lambda$, and $\cO_-$ of odd functions. In other words we may
decompose any regular function $f(\lambda)$ upstairs as
\be f(\lambda)\ =\ f_+(z) + \lambda\, f_-(z) \ee
Thus $E$ is generated by $m_1 = 1$ and $m_2 = \lambda$. To complete
the description, we must specify the action of $\lambda$, which is
clearly given by
\be\label{HiggsRamification} \lambda \cdot m_1\ =\ m_2, \qquad
\lambda \cdot m_2\ =\ z\, m_1 \ee
Using (\ref{HiggsAction}) we can then read off the Higgs field,
which is given by
\be \Phi\ =\ \left(
  \begin{array}{cc}
    0 & 1 \\
    z & 0 \\
  \end{array}
\right) \ee
The spectral equation reproduces $\det(\Psi) = \lambda^2 - z = 0$,
as expected.

At $z=0$, equation (\ref{HiggsRamification}) reduces to
\be \lambda \cdot m_1\ =\ m_2, \qquad \lambda \cdot m_2\ =\ 0\ee
This is precisely the structure sheaf of a double point:
\be \cO_{2p} \ = \ {\bf C}[\lambda]/\vev{\lambda^2 = 0} \ee
Although it should be clear by now, let us also check that this can
be recovered as the cokernel of $\Psi$, as discussed in subsection
\ref{HiggsvsSpectral}. At $z=0$ we have
\be \Psi\ =\ \lambda I - \Phi\ =\
 \left(
  \begin{array}{cc}
    \lambda & -1 \\
    0 & \lambda  \\
  \end{array}
\right) \ee
The image of $\Psi$ consists of pairs $(a(z)\lambda - b(z),
b(z)\lambda)$, where $a(z)$ and $b(z)$ are arbitrary polynomials in
$z$. The cokernel is therefore generated by
\be m_1\ =\ (0,1), \qquad m_2\ =\ (1,0) \ee
subject to the relations
\be\label{FatRelations} \lambda\cdot m_1\ \simeq\ m_2, \qquad
\lambda\cdot m_2\ \simeq\ 0 \ee
as required. By comparison, we also consider the cokernel of
\be \Psi \ = \ \left(
  \begin{array}{cc}
    \lambda & 0 \\
    0 & \lambda  \\
  \end{array}
\right) \ee
which corresponds to $\Phi = 0$. It is generated by the same $m_1$
and $m_2$, but instead it is subject to the relations
\be\label{ZeroRelations} \lambda \cdot m_1\ \simeq\ 0, \qquad
\lambda\cdot m_2\ \simeq\ 0 \ee
Clearly this is isomorphic to $\cO_p \oplus \cO_p$. This is a
perfectly legitimate sheaf on $2p$, it just differs from the
structure sheaf $\cO_{2p}$.

Let us consider one final perspective, which will be very useful
when we get to heterotic models. Note that the relations
(\ref{FatRelations}) are equivalent to saying that we have an
extension sequence
\be
 0 \ \to \cO_p \ \mathop{\to}^j\ \cO_{2p} \ \mathop{\to}^r \ \cO_p \
 \ \to \ 0
\ee
which does not split over ${\bf C}[\lambda]$. Here the `restriction
map' $r$ sets $\lambda \to 0$, i.e. $r(c_1 + \lambda c_2) = c_1$,
whereas $j(c_2) = \lambda\, c_2$. On the other hand, the relations
(\ref{ZeroRelations}) correspond to an exact sequence
\be
 0 \ \to \cO_p \ \mathop{\to}^j\ \cO_{p}\oplus \cO_p \ \mathop{\to}^r \ \cO_p
 \ \to \ 0
\ee
which does split.

One can easily generalize this discussion to fat points with length
greater than two, given by $\lambda^n = 0$. We leave this as an
exercise.

\newsubsection{Intersecting configurations}

Consider an intersecting configuration of two holomorphic cycles
$D_1,D_2$ in a Calabi-Yau three-fold $X$, and holomorphic line
bundles $L_1,L_2$ on each of them. It was shown in
\cite{Donagi:2010pd} that this data is not a complete description of
the configuration. In general, configurations which are reducible or
non-reduced are glued together by a gluing map, which should be
meromorphic in $B$-model-like settings. Therefore in addition, one
has to specify how the line bundles are glued along the intersection
$\Sigma=D_1 \cap D_2$. This gluing data is given by a meromorphic
section $f$ of $L_1^\vee \otimes L_2|_\Sigma$. It is usually
implicitly assumed that this gluing morphism vanishes. For most of
this paper, we will be interested in configurations where it does
not vanish.

When the gluing morphism vanishes, the massless spectrum (i.e. the
infinitesimal deformations) of open strings stretching between $D_1$
and $D_2$ can be computed as
\be\label{Ext1toHom} \Ext^1(i_{1*} L_1,i_{2*}L_2)\ \simeq\
\Hom_{D_1\cap D_2}(L_1,L_2 \otimes K_1) \ee
This looks very much like the gluing morphism above, except there is
a discrepancy involving the canonical bundle $K_{D_1}$.

The relation between the two was clarified in \cite{Donagi:2010pd}.
Let us instead start with a configuration ${\cal L}$ on $D_1\cup
D_2$, with restriction maps $r_1:{\cal L}\to L_1$ and $r_2:{\cal
L}\to \widetilde L_2$, where $L_1$ and $\widetilde L_2$ are line
bundles supported on $D_1$ and $D_2$ respectively.  We assume the
gluing morphism $f:L_1|_\Sigma \to \widetilde L_2|_\Sigma$ is
non-vanishing and holomorphic. Denote by $w$ a coordinate along the
intersection. On a small open set in $D_1 \cap D_2$, a local
section $p_1(w)$ of $L_1|_\Sigma$ can be lifted to a local section
of ${\cal L}|_\Sigma$:
\be\label{GluingSections} (p_1(w), f(w) p_1(w)) \ee
Now as we take the limit $f \to 0$, we see that local sections are
necessarily vanishing in the second argument. Thus the line bundles
we end up with in the limit are not $L_1$ and $\widetilde L_2$, but
instead $L_1$ and $L_2 \equiv \widetilde L_2 \otimes
\cO_{D_2}(-\Sigma)$. The massless modes of open strings stretching
between these two branes are given by
\be \Hom_{D_1\cap D_2}(L_1, L_2 \otimes K_1)\ =\ \Hom_{D_1\cap
D_2}(L_1,\widetilde L_2) \ee
Therefore deforming by this zero mode corresponds
precisely to turning on the gluing morphism on the brane
intersection. In particular, the support of the branes is unchanged,
so it does not correspond to recombining the intersecting branes
into a smooth irreducible configuration.

Alternatively, we can examine this from the point of view of the
Higgs bundle. Let us consider a Higgs field of the form
\be\label{OffDiagonalPhi} \Phi \ = \ \left(
  \begin{array}{cc}
     z/2 & f(w) \\
    0 & - z/2 \\
  \end{array}
\right) \ee
Let us also define $x \equiv \lambda - z/2$ and $y
\equiv\lambda+z/2$. Then the cokernel of $\lambda I -\Phi$ is the
sheaf ${\cal L}$ generated by $m_1,m_2$ such that
\be x m_2 - f m_1=0, \qquad y m_1 = 0 \ee
Now let us project this on ${\bf C}[x,w]\vev{m_1}$, which is a
sheaf that we will call $i_{1*}L_1$. It is easy to see that the
natural map ${\cal L} \to i_{1*}L_1$ is onto, and the kernel is
given by ${\bf C}[y,w]\vev{m_2}$, which we denote by $i_{2*}L_2$.
So the spectral sheaf is also described by the non-trivial
extension sequence
\be 0\ \to \ i_{2*}L_2 \ \to \ {\cal L} \ \to \  i_{1*}L_1 \ \to \
0 \ee
Although we made the argument on an open set, it holds on every
open set and therefore it is global. Now note that due to the
relation $x m_2 - f m_1=0$ the gluing morphism $f$ does not take
sections of $L_1|_\Sigma$ to sections of $L_2|_\Sigma$, but rather
to sections of $L_2\otimes \cO(\Sigma)|_\Sigma= L_2 \otimes
K_1|_\Sigma$. So we conclude that $f$ lives in (\ref{Ext1toHom}).

There are many equivalent ways to reach the same conclusions. Let us
discuss the point of view of the Higgs bundle a bit more. We can
engineer the brane intersection as an $SU(2)$ Higgs bundle over
${\bf C}^2$, parametrized by $(z,w)$, and with Higgs field
\be \Phi(z) \ = \ z  \ T_3\, dz\, dw, \qquad A^{0,1} =0
\ee
independent of $w$. Here we use the following notation for the
$SU(2)$ generators:
\be T^3 = \half\left(
  \begin{array}{cc}
    1 & 0 \\
    0 & -1 \\
  \end{array}
\right) , \qquad T^+ = \left(
  \begin{array}{cc}
    0 & 1 \\
    0 & 0 \\
  \end{array}
\right) , \qquad T^- = \left(
  \begin{array}{cc}
    0 & 0 \\
    1 & 0 \\
  \end{array}
\right) \ee
The equation for the spectral cover is
\be \det(\lambda I - \Phi)\  = \ \left(\lambda - {z\over
2}\right)\left(\lambda + {z\over 2}\right)\ =\ 0, \ee
which corresponds to a reducible configuration intersecting over
$z=0$. In \cite{Donagi:2008ca,Beasley:2008dc} it was shown that
there are localized zero modes
\be\label{LocalZeroMode} \delta A^{0,1} \ = \ e^{-z\bar z}\, T^+\,
d\bar z, \qquad \delta \Phi^{2,0} \ = \ e^{-z\bar z}\, T^+\, dz\,
dw\ee
To see the effect of such a deformation on the support of the
sheaves, we simply consider the spectral cover for the perturbed
Higgs field $\Phi + \epsilon \delta \Phi$. Clearly the equation for
the spectral cover is unchanged, so we see that the (holomorphic)
support is still reducible after turning on a VEV for this mode.

We can connect this to the previous point of view by applying
complex gauge transformation. Consider an infinitesimal
transformation with parameter
\be\label{gauge2} \lambda(z)\ =\ {1\over z}(1 - e^{-z\bar z})\, T^+
\ee
Applying this to (\ref{LocalZeroMode}), we find that we can express
the zero mode as
\be \delta A^{0,1}\ =\ 0, \qquad \delta \Phi\ =\ T^+\, dz\, dw\ee
Deforming by this zero mode yields
\be A^{0,1}\ =\ 0, \qquad \Phi \ = \
  \half \left(
  \begin{array}{cc}
    z & 1 \\
    0 & -z \\
  \end{array}
\right)\, dz\, dw \ee
This is manifestly holomorphic, and agrees with the algebraic
description we had earlier in (\ref{OffDiagonalPhi}), with the
off-diagonal generator corresponding to the gluing VEV. The value of
the off-diagonal generator is irrelevant away from $z=0$ because
$\Phi$ is diagonalizable there. We could also have applied a gauge
transformation with parameter
\be \lambda(z) \ = \ {1\over z}(e^{-z\bar z/m} -e^{-z\bar z})\,T^+
\ee
and take the limit $m \to 0$. Then we end up with the current
\be \delta A^{0,1} \to -\pi \delta(z)\, T^+\, d\bar z, \qquad \delta
\Phi \to 0 \ee
which is supported at $z=0$ but not holomorphic.

To summarize, when the branes intersect we have two inequivalent
choices for the Higgs field. The conventional choice is a vanishing
Higgs field. In terms of the spectral data, this corresponds to zero
gluing VEV along the intersection, in other words the spectral sheaf
looks like the rank two bundle $\cO \oplus \cO$ over the
intersection. The second possibility is a rank one Higgs field,
equivalent to a two-by-two Jordan block. In terms of the spectral
data, this corresponds to non-vanishing gluing VEV. In this case,
the spectral sheaf looks like the structure sheaf of a non-reduced
scheme over the intersection, as the equation for the spectral cover
over $z=0$ is given by $\lambda^2 = 0$. The second possibility is
actually simpler and more generic, for instance the simplest
possible sheaf on the reducible configuration is the structure sheaf
which is of the second type. We also still have to solve the
$D$-terms. This is discussed in section \ref{Dterms}.

It is easy to engineer both types of configurations as a
degeneration of a line bundle on a smooth irreducible configuration.
Let us consider a $U(2)$ Higgs bundle with Higgs VEV
\be
\Phi \ = \
  \half \left(
  \begin{array}{cc}
    z & 1 \\
    \epsilon & -z \\
  \end{array}
\right)
\ee
The spectral cover is given by
\be \left(\lambda - {z\over 2}\right)\left(\lambda + {z\over
2}\right)\  - {\epsilon\over 4}\ =\ 0 \ee
In the limit $\epsilon \to 0$, we end up with a reducible configuration
with non-zero gluing VEV. We may also consider a $U(2)$ Higgs bundle with Higgs VEV
\be\label{BFamily} \Phi \ = \
  \half \left(
  \begin{array}{cc}
    z & \delta \\
    \delta & -z \\
  \end{array}
\right)
\ee
This has exactly the same spectral cover, but in the limit $\delta
\to 0$ we end up with a reducible configuration with zero gluing
VEV. Note that in this case we effectively need an extra tuning to
set the gluing VEV to zero, so this is less generic.

Note also that the existence of the family (\ref{BFamily}) of
smoothing deformations is perfectly consistent with our picture of
brane recombination. Essentially it corresponds to turning on a VEV
of the form $\left< \right.\! Q\tilde Q\! \left. \right>$, where $Q$
and $\tilde Q$ are massless modes with opposite $U(1)$ charges. The
deformations with non-zero gluing VEV on the other hand have either
$Q=0$ or $\tilde Q=0$, and require a non-zero Fayet-Iliopoulos
parameter in order to satisfy the $D$-terms. In principle, one can
consider both of these deformations. When embedded in more
complicated set-ups however, turning on a VEV of the form $\left<
\right.\! Q\tilde Q\! \left. \right>$ is often forbidden by other
terms in the superpotential, and only the gluing VEV deformation is
available.

Let us also briefly discuss $A$-branes. This needs more
investigation, and our remarks will be more tentative.

If we are given intersecting Lagrangian branes, then once again we
have to decide what to do with the line bundle at the intersection.
We could glue the line bundles of the irreducible components at the
intersection using a gluing morphism, and we expect that this
corresponds to turning on a VEV for a chiral field localized at the
intersection, because the gluing morphism is clearly localized
there.

We can also discuss this in the language of real Higgs bundles
introduced in \cite{Pantev:2009de}. The gauge and Higgs field on a
real manifold $Q_3$ combine into a complexified connection, and the
$F$-terms say that this connection is flat. The $D$-terms yield an
equation for the hermitian metric, which splits the complex
connection into its anti-hermitian part $A$ and its hermitian part
$i\phi$. Generically one has $[\phi,\phi]\not=0$, but we can also
split the complex connection into a pair $(A,\phi)$ such that
$[\phi,\phi]=0$ almost everywhere on $Q_3$. Then we can diagonalize
and extract the spectral data, which can be represented as a
Lagrangian submanifold of $T^*Q_3$ with a flat unitary connection.
(Here as in \cite{Pantev:2009de} we assumed that the structure group
is reductive. When this is not the case, this picture should be
slightly generalized, see below).

 Let us denote by $f$ a harmonic
function on ${\bf R}^3$ with the flat metric. In fact we will take
$f = \half \sum_{i=1}^3 p_i x_i^2$ with $p_1 + p_2 + p_3=0$ and
$p_1,p_2>0$. Then we can describe a brane intersection by an $SU(2)$
Higgs bundle configuration of the form
\be i\phi\ =\ -\, df \, T^3 , \qquad A\ =\ 0\ee
In \cite{Pantev:2009de} we actually used a $U(2)$ Higgs bundle, but
this is not a material difference. The linearized version of the
$F$-terms is $d_{\bf A} \delta {\bf A} = 0$ where ${\bf A} = A + i
\phi$. We found the following localized solution at the intersection
(also satisfying the $D$-terms) \cite{Pantev:2009de}:
\be\label{ABraneLocSolution} \delta A + i \delta \phi \ = \
e^{-\half p_1 x_1^2 - \half p_2 x_2^2 + \half p_3 x_3^2}\,dx_3\, T^+
\ee
If we perturb by this solution, then we find
$[\phi_\epsilon,\phi_\epsilon]\not = 0$, where $\phi_\epsilon = \phi
+ \epsilon \delta \phi$. So although the intersection is in some
sense smoothed out, this does not yield a Lagrangian submanifold
with flat connection, but rather a kind of fat object. (The harmonic
metric, which actually determines the decomposition of ${\bf A}$
into a higgs field $\phi$ and gauge field $A$, is also changed, but
the decomposition in (\ref{ABraneLocSolution}) into $\delta A$ and
$\delta \phi$ should be valid to first order in $\epsilon$).

Thus now we appear to have at least two candidate deformations
corresponding to turning on a VEV for the chiral field at the
intersection. The second deformation however did not yield a
spectral cover. To get an analogy with what we did for $B$-branes,
we need an `abelianized' representative, i.e. we want to split ${\bf
A}_\epsilon$ into a pair $(A,\phi)$ such that
$[\phi_\epsilon,\phi_\epsilon]=0$ generically. Such a representative
does correspond to a Lagrangian submanifold with flat connection,
even when the harmonic representative does not. (Such a
representative would not be unique, since any Lagrangian related by
a normalizable hamiltonian deformation is still equivalent at the
level of $F$-terms.)

A connection with a non-reduced structure group cannot be
diagonalized. But we can decompose it in a semi-simple and a
nilpotent part, and take a sequence of complexified gauge
transformations such that the connection approaches the semi-simple
one. The semi-simple connection describes a Lagrangian brane with
unitary flat connection as usual. The original connection can then
be represented by this Lagrangian brane, except we have a non-zero
upper triangular part in the flat connection on the brane. This is
analogous to working with $S$-equivalence classes for bundles on
elliptic fibrations.

This association is easily done for our perturbed connection, as it
is already in upper-triangular form. The semi-simple part is simply
our original unperturbed solution. Thus we would like to propose
that the abelianized representative for our harmonic solution is
given by the original intersecting Lagrangian brane configuration,
but with a modified flat connection whose semi-simple part is
unitary. Equivalently this intersecting configuration has a non-zero
gluing VEV, given by a section of $L_1^\vee \otimes L_2|_p$ (i.e. a
single complex number) where $p$ is the point where the branes
intersect and $L_1, L_2$ are the flat $U(1)$-bundles on the two
components.

Our picture is also supported by results on mirror symmetry. It is
known that the category of $A$-branes should be extended to include
configurations of Lagrangian branes with flat connections that are
not quite unitary, but have monodromies with eigenvalues of
unit modulus \cite{PolisZaslow}.%
\footnote{The mnemonic is ``fat slags'' according to R.~Thomas.}
This allows for the possibility of Jordan block structure and is
precisely what we described above. In \cite{PolisZaslow} this Jordan
structure appeared along the whole Lagrangian brane, and in our case
essentially only at a point, but this is not a material difference.
Note also that turning on the gluing VEV would affect the morphisms
in the Fukaya category (discussed in section 3.2 of
\cite{PolisZaslow}) exactly as expected from turning on a VEV in the
superpotential.

The above picture does not exclude the existence of smoothing
deformations, and indeed Joyce has studied such examples
\cite{Joyce:1999tz}. The question however is whether the first order
infinitesimal deformations give rise to such a smoothing, and we
seem to find this is not the case. In fact in Joyce's picture, using
results of \cite{AkahoJoyce}, small deformations by modes on the
intersection should only deform the bounding cochain and thus also
leave the underlying Lagrangian submanifolds intact
\cite{JoyceCorrespondence}. This can probably also be understood by
thinking about intersecting branes in a hyperk\"ahler set-up,
because then $A$-branes and $B$-branes are related by a
hyperk\"ahler rotation.

One should also take into account normalizability. Let us consider
again the intersecting $B$-branes given by $xy=0$. From the point of
view of the branch parametrized by $x$, the smoothing mode is of the
form
\be \psi \ \sim \ {\epsilon\over x}\, dx \ee
If the hermitian metric approaches a constant for large $x$ in the
same frame in which the smoothing mode is given as above, as seems
reasonable, then the norm diverges as
\be \int^\Lambda rdr\, {1\over r^2} \ \sim \ \log \Lambda \ee
and so we could not ascribe the smoothing deformation purely to the
modes living at the intersection. On the other hand, the localized
modes we found in the field theory description have exponential
fall-off, and so are normalizable. We do not quite understand how to
reconcile this. Perhaps perturbing by $Q$ and $\tilde Q$
simultaneously is indeed not normalizable. At least this would seem
consistent with the fact that when embedded in more complicated
set-ups, integrating out KK modes typically leads to superpotential
terms of the form $W \sim (Q \tilde Q)^n$, which lifts the flat
direction for $Q \tilde Q$.

There is still a sense in which the intersection is smoothed out for
finite gluing VEV. Although the $F$-term data was completely
localized at $xy=0$, the solution to the $D$-terms has $[\Phi,
\Phi^\dagger] \not = 0$. The eigenvalues of ${\rm Re}(\Phi)$ and
${\rm Im}(\Phi)$ (with respect to the hermitian-Einstein metric) can
be identified with the position of the brane, at least in
perturbative type II. Since $[\Phi, \Phi^\dagger] \not = 0$ for
finite gluing VEV, the brane intersection is fattened and not
sharply localized. This is however a $D$-term effect, distinct from
the smoothing deformation taking $xy=0$ to $xy=\epsilon$ which is an
$F$-term effect. Our picture for intersecting $A$-branes with
non-zero gluing VEV has the same properties. We expect that this is
also the general picture for arbitrary intersecting brane
configurations: an expectation value for scalar fields at the
intersection with the same sign of the $U(1)$ charge corresponds to
a fattening deformation, and an expectation value for scalar fields
with opposite sign of the $U(1)$ charge corresponds to a smoothing
deformation.

We will discuss below how to compute the spectrum when the gluing
morphism is non-vanishing, but let us first discuss a further
generalization.

\newsubsection{Non-reduced configurations}

\subseclabel{NonReduced}

A second type of reducible brane is a configuration where the
divisor $D$ has some multiplicity. Such configurations are said to
be non-reduced schemes. As we will review later, the Fourier-Mukai
transforms of some of the most well-known heterotic bundles are
configurations of this type. Locally (i.e. fiberwise), these are
exactly the same structures that we saw arising at the ramification
locus and at brane intersections. A sheaf on a non-reduced scheme
may correspond to a smooth vector bundle localized on the support.
But one may also get sheaves that are non-trivial on the
infinitesimal neighbourhoods of $D$, in the sense that the
restriction map to the associated reduced scheme has a non-zero
kernel. Sheaves of this type were introduced in the $F$-theory
context in \cite{Aspinwall:1998he} and in the IIb context in
\cite{Donagi:2003hh}. They were studied in the context of mirror
symmetry in \cite{PolisZaslow}.

For simplicity again we first consider the case $R = 2D$, given by an
equation $\lambda^2 = 0$. Locally at
a generic point on $D$, this just reduces to the discussion of fat
points in section \ref{FatPoints}. Namely there are two natural rank
one sheaves, $\cO_{2D}|_p$ and $\cO_D \oplus \cO_D|_p$,
corresponding to Higgs fields of the form
\be \Phi = \left(
             \begin{array}{cc}
               0 & 1 \\
               0 & 0 \\
             \end{array}
           \right), \qquad
\Phi = \left(
             \begin{array}{cc}
               0 & 0 \\
               0 & 0 \\
             \end{array}
           \right)
\ee
respectively. Therefore here we will discuss the new issues that
arise in the general case. Then we have to consider situations where
the Higgs field vanishes or blows up over some curve in $D$.

Let us first consider a configuration where the Higgs field vanishes
along some curve $\Sigma$ in $D$, i.e. we have
\be \Phi\ =\ \left(
             \begin{array}{cc}
               0 & f \\
               0 & 0 \\
             \end{array}
           \right)
\ee
with $\Sigma = \{f=0\}$. We would like to establish the following
short exact sequence \cite{DEL}
\be\label{DELSES} 0\ \to\ \cO_D \ \mathop{\to}^j\ {\cal L}\ \mathop{\to}^r\
i_*K_D(-\Sigma)\ \to\ 0 \ee
Since the main new effect is that $f$ may vanish, we will focus on a
neigbourhood of a zero of $f$. Near such a zero, we can approximate
$f \sim  x$ where $x$ is a coordinate on $D$. Then ${\cal L}$ is generated
over the ring ${\bf C}[x,\lambda]$ by two generators
$m_1,m_2$, which are subject to the relations $\lambda m_2 = 0$ and $\lambda m_1 = x m_2$. In
other words we have
\be
{\cal L} \ = \ {\bf C}[x,\lambda]\vev{m_1,m_2}/(\lambda m_1 - x m_2, \lambda m_2)
\ee
Now the restriction map is given by setting $\lambda \to 0$, i.e.
\ba
\tilde r : {\cal L} \to {\cal L} \otimes {\bf C}[x,\lambda]/(\lambda) &=&
{\bf C}[x]\vev{m_1,m_2}/(x\, m_2) \eol
&=& {\bf C}[x]\vev{m_1} \oplus {\bf C}[x]\vev{m_2}/(x\, m_2)
\ea
Thus we get two pieces under the restriction. The first piece,
${\bf C}[x]\vev{m_1} $ is just a line bundle on $D$ which we will
denote by $L$, but the second piece is a torsion sheaf. We define a
new restriction map $r$ to be given by $\tilde r$ and them modding
out by the torsion, i.e. projecting on the first piece. So we have
\be
\  \ {\cal L}\ \mathop{\longrightarrow}^{r}\ i_* L\ \to \ 0
\ee
Now we need to find the kernel of $r$.

Let us consider a general section of ${\cal L}$, which is of the form
\be
(a_0 + a_1 \lambda + \ldots)\, m_1 + (b_0 + b_1 \lambda + \ldots)\, m_2
\ee
Under the restriction map $r$ this gets mapped to $a_0 m_1$. So the
kernel of $r$ is generated by sections of the form
\be
(a_1 \lambda) m_1 + b_0 m_2 \ = \ \left( {b_0\over x} + a_1 \right) \lambda \, m_1
\ee
where we used the relation $\lambda m_1 = x m_2$. Now $b_0/x + a_1$
generates $\cO(\Sigma)$, $\lambda$ generates $K_{D}^{-1}$, and
$m_1$ generates $L$. Thus the kernel of $r$ is identified as
\be
\cO(\Sigma) \otimes K_{D}^{-1} \otimes L
\ee
on $D$. Again we can make this argument on every open set and thus
it is global. The resulting sequence is therefore given by
\be\label{NRExtSequence} 0\ \to\ i_*(L(\Sigma) \otimes K^{-1})\
\mathop{\to}^j\ {\cal L}\ \mathop{\to}^r\ i_*L\ \to\ 0 \ee
which is equivalent to (\ref{DELSES}), as we wished to show.

Alternatively we can derive this sequence from the point of view of
the Higgs bundle. Suppose that $E$ is the sum of two line bundles,
$L_1$ and $L_2$. To get an irreducible object ${\cal L}$ we want to
turn on the off-diagonal component of the Higgs field. This
off-diagonal component $\Phi_{12} = f$ is a section of
\be L_1^\vee \otimes L_2 \otimes K \ee
Since $f$ is a section of $\cO(\Sigma)$ for some $\Sigma$, we see
that ${\cal L}$ is an extension
\be 0\ \to\ L_2\ \to\ {\cal L}\ \to\ L_1\ \to\ 0 \ee
where $L_2 = L_1(\Sigma)\otimes K^{-1}$.

Let us take a closer look at the extension class. We have
\be\label{NRExt1} \Ext^1(i_*L, i_*L(\Sigma) \otimes K^{-1}) \ \simeq
\ H^0(S,\cO_S(\Sigma)) \oplus H^1(S,K^{-1}(\Sigma)) \ee
We first interpret the first type of deformation in (\ref{NRExt1}).
Since $\Sigma$ is an effective divisor, there exists a section
vanishing at $\Sigma$, which we identify with $f(z)$ above. We can
interpret $f(z)$ as the gluing morphism, the off-diagonal generator
relating the zeroth and first order neighbourhoods. When it
vanishes, the sequence (\ref{NRExtSequence}) splits.

What about the remaining extension classes in (\ref{NRExt1}),
assuming they exist? They clearly correspond to changing the two
line bundles into a non-abelian rank two gauge bundle on $S$, i.e.
the traditional deformation corresponding to the extension sequence
on $S$:
\be 0 \ \to \ L(\Sigma) \otimes K^{-1} \ \to\ {\cal V} \ \to \ L \
\to \ 0 \ee
where ${\cal V}$ is a rank two bundle on $S$. It is satisfying to
see the two different types of deformation, the nilpotent Higgs VEV
yielding ${\cal L}$ and the non-abelian bundle deformation yielding
$i_*{\cal V}$, appear naturally from the $\Ext^1$.

If we have a Higgs field with larger Jordan blocks, then we can
iterate this construction. Consider for instance a Jordan block of
the form
\be \Phi\ = \
\left(
  \begin{array}{ccc}
    0 & f & 0 \\
    0 & 0 & g \\
    0 & 0 & 0 \\
  \end{array}
\right) \ee
This yields the relations
\be \lambda\cdot m_1\ =\ f m_2, \qquad \lambda\cdot m_2\ =\ g m_3,
\qquad \lambda\cdot m_3 = 0 \ee
We can first restrict this to the second order neighbourhood by
setting $\lambda^2 \to 0$ but $\lambda \not = 0$. Then we get a
natural projection to ${\bf C}[x,\lambda]\vev{m_1,m_2}/(\lambda m_1
- f m_2, \lambda m_2)$, which we denote as ${\cal L}_2$, and a
kernel which we can take to be $\cO_D$. Thus we have a short exact
sequence
\be 0 \ \to \cO_D \ \mathop{\to}^{j_2}\ {\cal L}\
\mathop{\to}^{r_2}\ {\cal L}_2 \ \to\ 0 \ee
Furthermore we recognize ${\cal L}_2$ to be the sheaf we treated
above. Then we have a second exact sequence
\be 0 \ \to \cO_D \ \mathop{\to}^{j_1}\ {\cal L}_2\
\mathop{\to}^{r_1}\ i_*K^2(-\Sigma) \ \to\ 0 \ee
Clearly we can set this up for any type of Higgs field $\Phi$. It
is also possible to create various in-between scenarios, eg. a rank
one sheaf on $3S$ which restricts to a rank two bundle on $S$.

We can easily give simple examples of the above types of
configuration. Suppose that $E$ is a sum of two line bundles, $E =
\cO(P) \oplus \cO(-P)$ for some divisor $P$ on a del Pezzo surface,
with zero Higgs field. As discussed in section \ref{Dterms}, this
configuration is unstable if the slopes of the two line bundles are
not equal, so the $D$-terms are not satisfied unless the slope of
$P$ vanishes. Now if $\delta \Phi_{12} \in H^0(S, \cO(2P)\otimes
K)$ is non-trivial then there are nearby configurations with a
nilpotent Higgs VEV. It is not hard to choose $P$ and the K\"ahler
class $J$ such that the resulting configuration is stable. We can
embed such non-reduced configurations in an $E_8$ Higgs bundle in
order to get new models. Some simple examples of $E_6$-models with
such non-reduced structure along the GUT brane are discussed in
section 2.2 of part II.

The next topic we want to discuss is possible poles for the Higgs
field. We consider a configuration of the form
\be \Phi \ \sim \
\left(
  \begin{array}{cc}
    0 & 1/z \\
    0 & 0 \\
  \end{array}
\right) \ee
Recall from section \ref{ParHiggs} that such a Higgs field should
be regarded as $K(D)$-valued, where in the notation of section
\ref{ParHiggs} the divisor $D$ on our surface is given by $z=0$.
The spectral equation seems to give $\lambda^2=0$, but something is
amiss as $|\Phi|^2$ diverges at $z = 0$. To get some idea about its
meaning, we slightly deform the Higgs field
\be \Phi \ \sim \ \left(
  \begin{array}{cc}
    0 & 1/z \\
    \epsilon & 0 \\
  \end{array}
\right) \ee
which should still be viewed as $K(D)$-valued. The spectral cover
is given by $\lambda^2 - \epsilon/z = 0$. This is the usual form of
spectral covers considered in \cite{Donagi:2009ra}. It clearly
corresponds to two sheets of the cover shooting off to infinity at
$z=0$, the eigenvalues growing as $\lambda= \pm \sqrt{\epsilon/z}$.
The cover is ramified at infinity over $z=0$. As a result, even
though we have two sheets going to infinity, the intersection
number with infinity is one.

If we now blindly take the limit $\epsilon \to 0$ above, we would
change the behaviour at infinity (in particular the intersection
number with infinity). Mathematically speaking, this is not a flat
family. Instead let us rewrite the spectral cover equation as
$z\lambda^2 - \epsilon = 0$, which for $z\not =0$ has the same
solutions. As $\epsilon \to 0$, we do not change the behaviour at
infinity, and the cover limits to $z \lambda^2 = 0$. That is, we get
the non-reduced scheme $\lambda^2=0$ away from $z=0$, and the
vertical fiber over $z=0$. In particular the intersection number
with infinity is still equal to one. Thus we interpret this as the
correct equation for the spectral cover.

In our previous work, we have avoided configurations where the
spectral cover has vertical components, because it would seem that
the $8d$ gauge theory description breaks down. This is perhaps too
pessimistic. As we saw above, the spectral cover for quasi-parabolic
Higgs bundles can have vertical components, and we can still study
wave functions that have a bounded $L^2$-norm.

On the other hand, there are also configurations with the same
equation for the spectral cover, and where the gauge theory
description really does break down. To see this, it helps to use
heterotic/$F$-theory duality. Consider a hermitian Yang-Mills bundle
$V$ on an elliptically fibered Calabi-Yau three-fold $\pi:Z \to
B_2$, in the limit that an instanton shrinks to zero size, and is
localized on a curve $D$ in the base. In the limit we end up with $V
\oplus \cO_D$, where $\cO_D$ is the structure sheaf of $D$, which
models some aspects of an $NS5$-brane wrapped on $D$. The
Fourier-Mukai transform of this is a spectral cover $C$ for $V$,
which is generically smooth, and a vertical fiber $\pi^*D$ which is
not glued to $C$. This is the small instanton transition. It is
non-perturbative and corresponds to a transition to a new branch,
with new degrees of freedom that cannot be seen in the $E_8$ gauge
theory description. It is a very singular point on the moduli space
of Higgs bundles. So in this case, the gauge theory description
really cannot be trusted. In the dual Calabi-Yau four-fold, it
corresponds to blowing up the base along $D$, which creates new
cycles along which the Ramond-Ramond four-form has additional zero
modes.

One can also study this system by introducing hypermultiplets on $D$
an studying the associated linear sigma model on $D$, as in sections
6.2 and 6.3 of \cite{Gukov:2006jk}. Here also one finds that the
quantum corrections become large and a new branch develops in the
limit of interest (called $P_0$ there). In the picture of
\cite{Gukov:2006jk}, on some slice of the configuration space these
large quantum corrections can be interpreted as instantons with
small action of the gauge theory on $S$ in the presence of a surface
operator on $D$, so the gauge theory on $S$ actually `knows' that it
is breaking down. In order to trust the gauge theory we should stay
away from this singular configuration.

To summarize, not all vertical fibers are created equal, and one
has to pay attention to the precise gauge theory configuration that
they correspond to. For more on this, see the section 4 of part II
on the $K3$ surface.

\newsubsection{Higgs bundles versus ALE fibrations}

Our discussion has focussed almost exclusively on Higgs bundles and
spectral covers. There is another correspondence which maps the
spectral cover to an elliptically fibered Calabi-Yau $Y_4$ with
$G$-flux, which yields the more traditional description of
$F$-theory vacua. One may wonder how the gluing morphism or a
nilpotent Higgs VEV appear in this picture, as naively there does
not seem to be room for gluing data there. In fact, in order to
write down an $F$-theory compactification we need to specify an
additional piece of data, namely a point on the intermediate
Jacobian:
\be {\cal J} \ = \ H^3(Y_4,{\bf R})/H^3(Y_4,{\bf Z}) \ee
This is usually ignored because for Calabi-Yau four-folds, the
intermediate Jacobian is often trivial. For the cases of interest
here however, it is in some sense not trivial, and this accounts for
the missing data.

To see this more precisely, it will be useful to first reconsider
the description of line bundles in the spectral cover picture,
because the Calabi-Yau fourfold picture is closely related to this.
Recall that holomorphic line bundles are classified by the Picard
group $H^1(\cO_C^*)$, and we have the long exact sequence
\be \to \ H^1(C,{\bf Z}) \ \to \ H^1(\cO_C) \ \to \  H^1(\cO_C^*) \
\to \ H^2(C,{\bf Z}) \ \to \ldots \ee
Thus to specify a line bundle, we need to specify the flux (the
first Chern class in $H^2(C,{\bf Z})$, and a point on the Jacobian
$H^1(\cO_C)/H^1(C,{\bf Z})$. In fact when the above sequence does
not split, we need additional information, but let us ignore that
here.

Let us consider a line bundle on a Riemann surface, say an elliptic
curve. The Jacobian is one dimensional and can be identified with
the dual of the elliptic curve. We can degenerate the elliptic curve
to a nodal curve, a ${\bf P}^1$ with two points identified. Line
bundles on ${\bf P}^1$ are completely classified by their flux, so
naively it seems the Jacobian has disappeared. This is not correct
because near the double point we can describe the curve by $xy=0$,
i.e. it looks like two intersecting curves. At $x=y=0$ we have to
specify the gluing morphism. Thus the Jacobian is still
one-dimensional in the limit. Similarly in the limit that a smooth
curve degenerates to a double curve (a `ribbon'), the Jacobian
degenerates but its dimension doesn't change.

We could also consider degenerating a degree two rational curve to
two intersecting degree one curves. Again we have an intersection
which looks like $xy=0$, and we have to specify a gluing morphism.
However we expect the Jacobian to be zero dimensional in this case,
since it is zero dimensional for the smooth curve. The reason this
works out is that the curve has become reducible and we get extra
automorphisms, so that any non-zero value of the gluing VEV can be
related to any other and hence any non-zero value of the gluing VEV
yields an isomorphic line bundle as far as complex structure is
concerned. After modding out by these automorphisms, and assuming
we fixed the flux, the moduli space appears to consist of three
points, where the gluing VEV is zero, finite or infinity. This is
not quite right because zero and infinity are in the closure of
finite gluing VEV. Rather, the moduli space consists of ${\bf
CP}^1$ modulo a ${\bf C}^*$-action. It is not a smooth space, but
rather a stack, i.e. roughly speaking a kind of scheme with an open
subset corresponding to finite gluing VEV, and the points with zero
and infinite VEV embedded as negative dimensional closed
subschemes.

These phenomena have a simple physical description in terms of the
Higgs mechanism, as explained in more detail in section
\ref{SimpsonStability}. For finite gluing VEV the would-be
$h^1(\cO_C)$ which corresponds to changing the gluing VEV is eaten
by a would-be generator of $h^0(\cO_C)$. However physically we also
have to split the deformation in a real part and an imaginary part.
The imaginary part becomes the longitudinal generator of a gauge
boson and the real part is lifted by a $D$-term potential. The
$D$-term potential contains a scale, set by the Fayet-Iliopoulos
parameter, which is a function of the K\"ahler moduli but not of the
complex structure moduli. Thus in contrast to the previous example,
different non-zero values of the gluing VEV yield isomorphic line
bundles as far as the complex structure is concerned, but they are
not the same physically, and this should be understood as a K\"ahler
modulus.

Situations like the above will arise in the context of
heterotic/$F$-theory duality in six dimensions. For
compactifications to four dimensions, we instead need to consider a
spectral surface in a Calabi-Yau three-fold. The case of spectral
surfaces (as opposed to spectral curves) is slightly different in
that there is a branch structure and the dimension of the moduli
space can be different on different branches. It is usually
comparable with the second situation, although we will see examples
with continuous moduli as well. Generic surfaces have
$h^1(\cO_C)=0$ and line bundles on them don't have moduli. However
when we degenerate them, the situation locally looks like that for
curves. The gluing VEV is part of the continuous data specifying
the spectral line bundle, so in a moral sense it should be
understood as defining a point on the `Jacobian' of the singular
spectral cover $C$. But the would-be generators of $h^1(\cO_C)$
corresponding to changing the gluing VEV are usually eaten by a
would-be generator of $h^0(\cO_C)$, or lifted by pairing with a
would-be generator of $h^{2}(\cO_C)$. In certain limits they may
appear in pairs. Thus in the reducible case the moduli space of the
spectral sheaf is often zero dimensional, and is not a smooth
space. But on certain branches, the gluing data may yield a
positive dimensional `Jacobian', like in the example of the
elliptic curve.

These statements have analogues in the elliptic Calabi-Yau picture
of $F$-theory, although there are important subtleties which we
discuss further below. The configuration of the three-form field
$C_3$ corresponds to a Deligne cohomology class. It is (roughly)
specified by a $G$-flux, where $G=dC_3$, and a point on the
intermediate Jacobian ${\cal J}$. In fact recall that the relation
between the spectral cover and the ALE-fibration is given by a
version of the cylinder mapping \cite{Curio:1998bva,Donagi:2009ra}.
The spectral cover determines the ALE fibration, and the spectral
sheaf determines a configuration for $C_3$. The Jacobian of $C$ and
the intermediate Jacobian of $Y_4$ are related by a cylinder map.
Again, this is a little loose because the moduli space may not even
be smooth, and looks nothing like an abelian variety, so we should
probably not call it a Jacobian. But at any rate we see that the
gluing data is not related to the complex structure of the
Calabi-Yau four-fold. Rather, it is part of the data needed to
specify a configuration for the three-form field $C_3$. For
example, an intersecting brane configuration of the form $xy=0$
gets mapped to a conifold singularity of the form $xy + zw = 0$,
and the message of the dictionary is that the physics depends on
the configuration of $C_3$ on this singularity. Similar remarks
apply to non-reduced configurations.

This leads to some interesting new issues in the study of
four-folds with flux. Using this dictionary, we can now resolve
several issues that previously looked very puzzling from the
$F$-theory/$7$-brane perspective, and fit it in the standard set-up
of geometric invariant theory. When we go to the $M$-theory
description on the resolved Calabi-Yau four-fold, we know that the
$D$-terms are given by $J \wedge G = 0$. As long as these equations
are valid, there are no stability walls. This might seem puzzling
because such walls do exist for example in the heterotic string,
which can arise as a small volume limit of $M$- or $F$-theory.
Accordingly it has been speculated that $11d$ supergravity just
sees one particular chamber in the moduli space of an $M$- or
$F$-theory compactification.

With the results in this paper, we can now see this chamber
structure more explicitly using a weakly coupled $8d$ gauge theory
description. In section \ref{Dterms} we will see that the VEVs of
the gluing data are set by Fayet-Iliopoulos terms, which would be
given by expressions of the form $\int G \wedge J \wedge \omega$ on
the (singular) four-fold. Thus the Higgs bundle picture is telling
us that in the deep $F$-theory regime where we can trust the $8d$
gauge theory description, but far from the regime where $11d$
supergravity is valid, the Fayet-Iliopoulos terms may be non-zero.
Hence we will argue that the traditional primitiveness condition $J
\wedge G=0$ should be generalized to a kind of stability condition,
coinciding with the stability condition for the Higgs bundle when
the $8d$ gauge theory description is valid, and that one does get a
chamber structure in the K\"ahler moduli space with walls of
marginal stability.

 \begin{figure}[t]
\begin{center}
            \scalebox{.4}{
               \includegraphics[width=\textwidth]{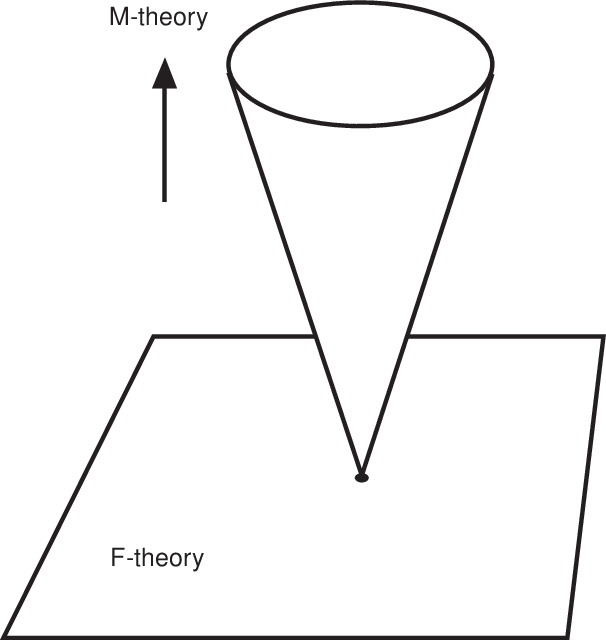}
               }
\end{center}
\vspace{-.5cm}
\begin{center}
\parbox{14cm}{\caption{ \it Picture of the branch structure. The cone
represents the $3d$ Coulomb branch, where one resolves the singularities
of the Calabi-Yau four-fold.
The plane represents a `non-abelian' $F$-theory branch where wrapped $M2$-branes have
condensed, eg. a branch with a non-zero Fayet-Iliopoulos parameter.
This branch is visible in the Higgs bundle description but not
in the Calabi-Yau four-fold description. The picture is schematic in
several respects, for example it is not guaranteed that
every $F$-theory branch is connected to an $M$-theory branch.}\label{MFBranches}}
\end{center}
 \end{figure}

We can further sharpen the claim that the $11d$ supergravity
approach is not giving the full picture of singular $F$-theory
compactifications. Recall that in this approach, one compactifies
on an extra circle to three dimensions. The four-dimensional vector
multiplet gains a pseudo-scalar upon compactification. Moving out
on the $3d$ Coulomb branch makes the non-abelian gauge bosons very
massive. In the dual $M$-theory picture, this scalar corresponds a
K\"ahler modulus $\delta J = t_X \omega^X$ where $\omega^X$ is the
$(1,1)$ form yielding a $U(1)_X$ gauge symmetry, $\delta C_3 = A_X
\wedge \omega^X$. Moving out on the Coulomb branch corresponds to
making a small resolution, and taking the size of the exceptional
cycles to be large, see figure \ref{MFBranches}. In the $M$-theory
picture, an $M2$-brane wrapped on a cycle $\alpha$ has a mass
proportional to
\be
\int_\alpha J\ \sim\  q_\alpha^X\, t_X,
\ee
where $q_\alpha^X =\int_\alpha \omega^X$ is its $U(1)_X$ charge, so
in this limit, we can quantize $M2$-branes wrapped on the
exceptional cycles.

Now we have seen that non-trivial configurations for $C_3$, such as
arising from gluing VEVs or non-trivial bundles on non-reduced components
of the spectral cover, can lift
vector multiplets, and therefore part or all of the $3d$ Coulomb
branch can get lifted. In particular, if a $U(1)_X$ gauge symmetry gets
lifted by $C_3$ (eg. when the Fayet-Iliopoulos term is non-zero) then $t_X$
is frozen at zero and there are cycles which cannot get resolved.%
\footnote{We note that the St\"uckelberg mechanism may also lift part of the Coulomb branch,
but there is an important difference. There masses are small and the mechanism can be seen
in $11d$ supergravity, whereas turning on gluing modes involves condensing non-perturbative BPS states.}
That is, the background value of the three-form field ${\sf C}_3$ can obstruct
the small resolution, and the soliton quantization approach is not applicable
to these interesting configurations, which as we shall see in part II
give rise to all kinds of interesting flavour structures. This is
analogous to the question of
whether one can go through a conifold transition: it does not depend
only on the geometry but also on the background fields, as they may
lift the light fields whose VEV controls the transition.

\newsubsection{Spectra of degenerate Higgs bundles}

\subseclabel{HiggsSpectra}

 Now we would like to understand how to
compute the spectra. As mentioned previously, these correspond to
the infinitesimal deformations and are computed by the
hypercohomology groups $\mathbb{H}^p({\cal E}^{\, \bullet})$ of the
Higgs bundle. On the other hand, the most concrete way of
constructing Higgs bundles is through the spectral data, so it would
be most convenient to compute directly with this data. The
hypercohomology groups can be directly computed in terms of the
spectral data:
\be \mathbb{H}^p({\cal E}^{\, \bullet}) \ = \ \Ext^p({\cal L},{\cal
L}) \ee
Similarly we
can compute the holomorphic couplings using Yoneda pairings. The
$D$-terms are discussed in section \ref{Dterms}.

The basic strategy for the computation of any $\Ext$ group is to
perform some kind of `resolution,' i.e. relate ${\cal L}$ to some
simpler sheaves, and then consider an associated long exact
sequence. We can intuitively understand this as expressing a brane
as a bound state, obtained by gluing simpler constituents together.
Let us see how this works for degenerate cases.

The sheaf ${\cal L}$ decomposes into several pieces, and we are
actually usually interested in computing $\Ext$-groups of the form
\be \Ext_X^p(E,{\cal L}) \ee
where $X$ is our Calabi-Yau three-fold. To do this, let us suppose
we can express ${\cal L}$ as an extension:
\be\label{StandardExtSeq} 0 \ \to \ B\ \to\ {\cal L} \ \to \ A \ \to
\ 0 \ee
To compute $\Ext_X^1(E, {\cal L})$ and $\Ext_X^1({\cal L},E) =
\Ext_X^2(E,{\cal L})^*$, we use the associated long exact sequence:
\be\label{TruncLongExSeq}
\begin{array}{ccccccccc}
  0 & \to & \Ext^1(E,B) & \to & \Ext^1(E,{\cal L})& \to & \Ext^1(E,A) & &
  \\[3mm]
   & \to  &\Ext^2(E,B) & \to & \Ext^2(E,{\cal L}) & \to
   & \Ext^2(E,A) & \to & 0
\end{array}
\ee %
In normal situations, the $\Ext^0$'s and $\Ext^3$'s all vanish,
which we have assumed above to simplify the long exact sequence.
This is not a limitation. If it is not satisfied, the story is much
the same as below, except some additional generators may get lifted
through the Higgs mechanism (which lifts $\Ext^0$ and $\Ext^1$
generators in pairs). But let us assume this is not needed here.
Then we find that $\Ext^1(E,{\cal L})$ is generated by $\Ext^1(E, A
\oplus B)$, except that some generators of $\Ext^1(E,A)$ may get
killed by the coboundary map.

The mathematics of the long exact sequence can be expressed in terms
of the effective Lagrangian of the brane system. In the brane bound
state picture, we have deformations involving the constituent branes
$A$ and $B$, i.e. we have chiral fields
\be X_1 \in \Ext^1(E,B),\quad  X_2 \in \Ext^1(B,E), \quad Y_1 \in
\Ext^1(E,A), \quad Y_2 \in \Ext^1(A,E) \ee
Now all the $X_p,Y_p$ may in principle descend to generators in
$\Ext^p(E,{\cal L})$. However, some $Y_1,X_2$ pairs may be lifted by
interactions. In fact the coboundary map is simply the Yoneda
pairing
\be \Ext^1(E,A) \times \Ext^1(A,B) \ \to \ \Ext^2(E,B) \ee
In other words, there are Yukawa couplings for the chiral fields
\be W \ \simeq \ Y_1 \, F_{glue}\,X_2 \ee
where $F_{glue}\in \Ext^1(A,B)$ is the extension class. So when the
gluing morphism $F_{glue}$ gets a VEV and we form the bound state
${\cal L}$, we see that the $X_1$ and $Y_2$ fields may pair up and
get a mass through their Yukawa couplings to $F_{glue}$. This is how
the lifting through the coboundary map translates to the effective
Lagrangian. The surviving chiral fields correspond to the
deformations in $\Ext^1(E,{\cal L})$ that we are after.

Note also that this is consistent with the charges under the extra
$U(1)$ symmetry that appears as the gluing map is turned off. Up to
an overall normalization, these charges are given by
\be Q(X_1) = -Q(X_2) = -Q(Y_1) = Q(Y_2) = +1, \qquad Q(F_{glue}) =
+2 \ee
In particular, the above Yukawa coupling is the only one allowed by
the symmetries.

If $E = {\cal L}$, then we can also resolve $E$ using a short exact
sequence, and get a second long exact sequence involving the first
argument of $\Ext$. Although the algebra gets more involved, it is
in principle straightforward.

Let us apply this to the degenerate configurations in this paper.
Consider first an intersecting configuration ${\cal L}$, with a
non-zero gluing VEV. The support of ${\cal L}$ consists of two
divisors $D_1$ and $D_2$, but the configuration should really be
thought of as a single brane, as only the center-of-mass $U(1)$
gauge symmetry is unbroken. Let us denote by $i_1$ the inclusion
$D_1 \hookrightarrow  X$, and similarly for $D_2$. Since the support
is reducible, we have natural restriction maps to each component.
Now suppose that the restriction $i_1^*{\cal L}\ =\ L_1$ is actually
a line bundle. Then we can express ${\cal L}$ as an extension on
$X$:
\be\label{RedShortSeq} 0 \ \to \ i_{2*}L_2(-\Sigma) \ \to\ {\cal L}
\ \to \ i_{1*}L_1 \ \to \ 0 \ee
The second map is restriction to $D_1$ and then pushing forward to
$X$. This is of the form (\ref{StandardExtSeq}), so we can apply the
discussion above. The extension class is given by a holomorphic map
in $\Hom_{\Sigma}(L_1,L_2)$. Similarly if the restriction to $D_2$
yields a line bundle, then we get an analogous extension sequence
with $1$ and $2$ reversed.

Now in general the restriction to $D_1$ does not yield a line
bundle, but a sheaf with torsion. We only know that there is a
birational isomorphism between $L_1|_\Sigma$ and $L_2|_\Sigma$.
Instead of working with a meromorphic map, an equivalent way to say
this is there is another line bundle $L_\Sigma$ on $\Sigma$, and a
pair of holomorphic maps in $\Hom_{\Sigma}(L_1|_\Sigma,L_\Sigma)$
and $\Hom_{\Sigma}(L_2|_\Sigma,L_\Sigma)$. Then we have the short
exact sequence
\be\label{RedFundShortSeq} 0\ \to\ {\cal L}\ \to\ i_{1*}L_1 \oplus
i_{2*}L_2\ \to\ i_{\Sigma *}L_\Sigma\ \to\ 0 \ee
In other words, ${\cal L}$ is what might be called a Hecke transform
of $ i_{1*}L_1 \oplus i_{2*}L_2$ along $\Sigma$. In this case we
need the full long exact sequence for $\Ext$, not just the truncated
version (\ref{TruncLongExSeq}), but the advantage is that it applies
generally.

Let us briefly check that (\ref{RedShortSeq}) is indeed a special case of
(\ref{RedFundShortSeq}). Then we assume that the map ${\cal L} \to i_{1*}L_1$ is onto.
 The sheaf ${\cal L}$ is locally generated by sections
$(s_1,s_2)$ such that $f_1 s_1 + f_2 s_2=0$ on the intersection. Now let us ask for
the kernel of the map ${\cal L} \to i_{1*}L_1$. It is generated by sections $s_2$ such that
$f_2 s_2|_\Sigma=0$, i.e. $s_2|_\Sigma = 0$ assuming $f_2$ does not vanish identically.
But this is the definition of $i_{2*}L_2(-\Sigma)$.

 Similarly we may consider the case that ${\cal
L}$ consist of a line bundle over a non-reduced surface. For the
simplest case where the Higgs field is a Jordan block of rank two,
we found the extension sequence
\be 0\ \to\ i_*(L(\Sigma) \otimes K^{-1})\ \mathop{\to}^i\ {\cal L}\
\mathop{\to}^r\ i_*L\ \to\ 0 \ee
When there are Jordan blocks of higher rank, as discussed we can
iterate this. This is again of the form (\ref{StandardExtSeq}), so
we temporarily replace ${\cal L}$ by $A \oplus B$, where $A = i_*L$
and $B = i_*(L(\Sigma) \otimes K^{-1})$, and then lift pairs of
deformations by turning on the extension class in the long exact
sequence.

In cases where we are already given some explicit representative of
the non-abelian holomorphic bundle $E$ and the Higgs field $\Phi$,
we can use the short exact sequence (\ref{CokerDef}) to find ${\cal
L}$ and then compute $\Ext$ groups. Probably it is then simplest to
use computer algebra.

\newsubsection{Chiral matter and the index}

\subseclabel{Index}

 In the previous subsection we explained the
tools to compute the matter content of the theory. The calculations
are in principle straightforward, and can even be carried out by
computer algebra systems like Macaulay2. However in order to get a
quick overview of a model, it is often sufficient to know only the
{\it net} amount of chiral matter. This can be computed much more
efficiently using the index theorem:
\be \chi(F,G)\ =\ \sum_{i=1}^3 (-1)^i\, \Ext^i(F,G)\ =\ \int_X {\bf
ch}(F^\vee){\bf ch}(G) \hat{\bf A}(TX) \ee
For the cases of interest, it is often true that $\Ext^0$ and
$\Ext^3$ vanish (no ghosts), and so $\chi$ reduces to the net amount
of chiral matter. These formulae make sense for the reducible and
non-reduced cases we are interested in. It applies when $F$ and $G$
are merely coherent sheaves (or even complexes thereof), and $X$ is
projective (see eg. \cite{ResDuality,FultonIntersectionThy}).

In the IIb context this formula can be understood from anomaly
inflow. In the $F$-theory context, the `charge vector' ${\bf
ch}(F){\bf A}(TX)^{1/2} \in H^{even}(X)$ is not  defined on the IIb
space-time, but on the auxiliary Calabi-Yau three-fold $X$. In this
context we actually only need part of the charge vector; it can be
related to the couplings of the NS two-form under
$F$-theory/heterotic duality, and is therefore again closely tied to
anomalies.

To find the Chern classes of more complicated sheaves, we can use
one of the fundamental properties of the Chern character. If we have
a short exact sequence
\be 0\ \to\ M'\ \to\ M\ \to\ M''\ \to\ 0 \ee
then
\be {\bf ch}(M)\ =\ {\bf ch}(M') + {\bf ch}(M'')\ee
The index formula also involves the dual, $F^\vee$. In good
situations, the dual is again a sheaf, for instance for a line
bundle on a smooth divisor in $X$ we have $(i_*L)^\vee = i_*(L^\vee
\otimes K_D)$. More generally, it is not possible to require that
the dual is another sheaf while preserving all the expected
properties, and the dual is instead given by a complex, $F^\vee =
RHom^\bullet(F,K_X)$ \cite{ResDuality}. Fortunately for our purposes
we only need the following property:
\be {\rm ch}_i(F^\vee)\ =\ (-1)^i {\rm ch}_i(F) \ee

Let us apply this to the cases considered in this paper. For a
vector bundle $L$ supported on a smooth divisor $D$ we have
\be
\begin{array}{cclcccl}
  {\rm ch}_0(i_*L) &=& 0 & \qquad & {\rm
ch}_2(i_*L)& =& i_{*}c_1(\hat L)  \\
  {\rm ch}_1(i_*L) &=& {\rm rk}(L)\,D & \qquad & {\rm  ch}_3(i_*L) &= &
i_*\left({\rm ch}_2(\hat L)+{1\over 24}{\rm rk}(L)c_1(K_D)^2\right)
\end{array}
\ee
where $\hat L = L \otimes K_D^{-1/2}$. The twisting by $c_1(K_D)$ is
familiar from the Freed-Witten anomaly, which says that the gauge
field really takes values in the bundle $L \otimes K_D^{-1/2}$ on
$D$, and so the flux is given by
\be {{\rm Tr}(F)\over 2\pi}\ =\ c_1(\hat L)\ =\ c_1(L) - \half \,
{\rm rk}(L)\, c_1(K_D) \ee
Using the above, it is very simple to reproduce the standard formula
for the net amount of matter localized on brane intersections or in
the bulk of a 7-brane, assuming no gluing morphisms are turned on.
But we can equally well do the degenerate configurations. For the
reducible case, we used the short exact sequence
\be 0\ \to\ i_{2*}L_2(-\Sigma)\ \to\ {\cal L}_1\ \to\ i_{1*}L_1\
\to\ 0 \ee
Therefore we find
\be\label{ReducibleChern}
\begin{array}{rcl}
  {\rm ch}_0({\cal L}_1)& = & 0 \\
  {\rm ch}_1({\cal L}_1) & = & D_1 +
D_2 \\
  {\rm ch}_2({\cal L}_1) & = & i_{1*}c_1(\hat L_1) +
i_{2*}c_1(\hat L_2) - i_{\Sigma *}\Sigma
\end{array}
\ee
Similarly, in the non-reduced case we had the sequence
\be 0\ \to\ i_*(L(\Sigma) \otimes K_D^{-1})\ \mathop{\to}^i\ {\cal
L}\ \mathop{\to}^r\ i_*L\ \to\ 0 \ee
and from this we find
\be\label{NonReducedChern}
\begin{array}{rcl}
  {\rm ch}_0({\cal L}_2)& = & 0 \\
  {\rm ch}_1({\cal L}_2) & = & 2D \\
  {\rm ch}_2({\cal L}_2) & = & 2\,
i_*c_1(\hat L) - i_*c_1(K_D)+ i_{\Sigma *} \Sigma
\end{array}
\ee
As a simple example, let us consider the reducible brane ${\cal
L}_1$, and intersect it with another brane $i_{3*}L_3$. We find
\be \chi({\cal L}_1,i_{3*}L_3)\ =\ \int_{D_1 \cdot D_3} \left[
c_1(\hat L_1) - c_1(\hat L_3)\right] + \int_{D_2 \cdot D_3} \left[
c_1(\hat L_2(-\Sigma)) - c_1(\hat L_3)\right] \ee
This is the conventional formula when we turn off the gluing VEV.
Indeed the index should not change under such a continuous
deformation.  It should be remembered however that when the gluing
morphism has both poles and zeroes, then it cannot be turned off
holomorphically, and we have to use (\ref{RedFundShortSeq}) instead.

Similarly, let us consider the intersection of the non-reduced brane
${\cal L}_2$ with $i_{3*}L_3$. Then we have
\be \chi({\cal L}_2,i_{3*}L_3)\ =\  \int_{D \cdot D_3} \left[
c_1(\hat L)-c_1(\hat L_3) \right]+ \left[ c_1(\hat
L(\Sigma))-c_1(K_D)-c_1(\hat L_3) \right]  \ee
as we would when the gluing data is turned off.

\newsubsection{Boundary CFT description}

We have described non-reduced schemes as configurations in
supersymmetric Yang-Mills theory. In the type II context, one
naturally asks if there is also a boundary CFT description. The
first thing to try is a free-field description. Normally we would
have
\be \del_1 X(\sigma)|_0\ =\ 0 \ee
and then we tensor with Chan-Paton indices. For non-reduced
configurations we want instead
\be \del_1 X^2(\sigma)|_0\ =\ 2 X \del_1 X|_0\ =\ 0 \ee
and further we want $\del_1 X(\sigma)|_0 \not = 0$, for otherwise we
reduce to the previous case. This is a non-linear condition on the
mode expansion. In some sense this indicates we are dealing with
true non-abelian configurations. Therefore it does not seem likely
that we can find a free-field description.

There are however other methods for constructing boundary CFTs. One
such description is the boundary linear sigma model
\cite{Hori:2000ck,Govindarajan:2000ef,Hellerman:2001bu}. It can be
developed largely in parallel with $(0,2)$ linear sigma models,
which we briefly review in section 4.3 of part II.

We will keep things extremely simple and only explain the main idea.
Apart from the $(2,2)$ chiral fields $X_i$ and vector multiplet in
the bulk, we consider boundary chiral fields $P$ and boundary Fermi
fields $\Lambda_a, \Gamma$. We have a boundary superpotential
\be \int dx^0\,d\theta\ \Gamma\, S(X_i) + \Lambda_a P
J^a(X_i)|_{\bar \theta=0} + h.c. \ee
The $\Lambda_a$ lead to Chan-Paton factors, and $\Gamma$ is designed
to pair up with bulk fermions normal to $S(x_i)=0$. In the large
volume regime, the massless modes of $\Lambda_a$ live in a bundle
$\tilde V$ on $X$ defined by a short exact sequence
\be
 0\ \to\ \tilde V\ \to \ \bigoplus_{a} \cO(q_a)\
  \mathop{\longrightarrow}^{J^a}\ \cO(q_0)\ \to\ 0
\ee
The effect of the first term of the boundary superpotential is then
to restrict the open string ends to $S(x)=0$, so that we end up with
$V = \tilde V|_{S=0}$. This basic construction can be extended in
several directions.

This allows us to construct CFT descriptions of non-reduced
configurations. For instance the structure sheaf $\cO_{2D}$ of a
non-reduced scheme $2D$ fits into the exact sequence
\be 0\ \to\ \cO_X(-2D)\ \to\ \cO_X\ \to\ \cO_{2D}\ \to\ 0 \ee
Taking the dual, this naturally fits in the boundary LSM description
above. Similarly, one can construct the structure sheaf $\cO_D$ of a
reducible divisor, by taking a section of $\cO_X(D)$ which is
factorizable. This configuration has a non-zero gluing VEV along the
intersection of the irreducible pieces. We expect that the linear
sigma model flows to a CFT only when the configuration is (Gieseker)
poly-stable.

\newpage

\newsection{The D-terms}

\seclabel{Dterms}

\newsubsection{The hermitian-Einstein metric and stability}

\subseclabel{hEinsteinMetric}

In previous sections, we studied the $F$-terms of the $8d$ gauge
theory. $F$-flatness is preserved under complexified gauge
transformations. Modulo such complexified gauge transformations, the
only invariant data in the $F$-terms is the spectral data.  We used
this extensively for writing down solutions for the $F$-term
equations, by writing down the spectral sheaf.

The $D$-terms for the $8d$ gauge theory compactified on a K\"ahler
surface $S$ are given by the following `hermitian-Einstein'
equation:
\be\label{DEquation} g^{i\bar j} F_{i\bar j} +
[\Phi^{2,0\dagger},\Phi^{2,0}]\ =\ -\sqrt{-1}\,\zeta I \ee
with $\zeta \simeq {\rm deg}(E)/(r\, {\rm vol}(S))$. Here we think
of the commutator as a $(0,0)$-form by contracting with the volume
form of $S$. Unlike the $F$-terms, the $D$-terms are not invariant
under the complexified gauge transformations. They require us to
choose a hermitian metric, or equivalently a reduction of the
complexified structure group to a compact subgroup.

It may be useful to briefly recall some aspects of connections on
holomorphic vector bundles \cite{GriffithsHarris}. A frame for $E$
over an open subset $U$ is a collection of sections $H =
\{e_1,..,e_r \}$ forming a basis for each fiber over $U$. With a
suitable choice of coordinates on the fiber, we can write the
hermitian metric in matrix notation as
\be h\ =\ H  H^\dagger \ee
where $H$ is a map from $S$ to $GL(n,{\bf C})/U(n)$. The frame is
said to be unitary if $h_{a\bar b} = \delta_{a\bar b}$. The frame
is said to be holomorphic if the $e_a$ are holomorphic maps from
$U$ to $E$.

If the structure group is to be $U(n)$ rather than $GL(n,{\bf C})$,
then the gauge covariant derivative must respect the hermitian
metric. In the unitary frame, this implies that $A^+ = - A$, where
the superscript denotes transpose and complex conjugation. In a
more general frame however, this implies that $A^\dagger = - h^{-1}
A^+ h$, i.e. the adjoint depends on the hermitian metric. Similarly
the adjoint $\Phi^\dagger$ corresponds to $h^{-1} \Phi^+ h^{}$.

We can further fix the connection by requiring the connection to be
compatible with the complex structure, i.e. the $(0,1)$ part of the
covariant derivative is given by $\delb$ (so $A^{0,1}=0$ in a
holomorphic frame). In such a frame, the zero modes and
superpotential are independent of the K\"ahler moduli, and
computations reduce to questions in complex geometry and algebraic
geometry. With this additional condition, we find that the
connection is uniquely determined as
\be A^{1,0}\ =\ -( \del h)h^{-1} \ee
in the holomorphic frame. This connection is sometimes referred to
as the Chern connection. We can switch back to a unitary frame by
performing a complex gauge transformation by $H$. In this unitary
frame the connections are given by
\be H^{-1}(\del + A^{1,0})H \ =\ \del - (\del
H^\dagger)H^{\dagger\,-1} , \qquad H^{-1} \delb H\ =\ \delb +
H^{-1}(\delb H) \ee
Thus assuming we have fixed the $F$-term data, we see that the
$D$-terms may be viewed as the following equation for the hermitian
metric $h$ on $E$:
\be g^{i\bar j} \del_{\bar j}(  \del_i h \, h^{-1})+
[h^{-1}\Phi^+h,\Phi] - \sqrt{-1}\,\zeta I\ =\ 0 \ee
The solution is usually called the hermitian-Einstein metric. To
distinguish it from the hermitian metric which arises as a special
case when $\Phi = 0$, we might also call it the hermitian
Yang-Mills-Higgs metric, but this terminology is perhaps too
lengthy.

The solution to the abelian part of this equation can be found by
making a conformal change in the metric, $h \to h e^f$ and solving
for $f$. The non-abelian equations are much harder to solve. We
could solve the $F$-terms by writing down suitable spectral data.
However this approach does not work for the $D$-terms, even in the
generic case where the eigenvalues are mutually distinct over an
open subset, and thus we can diagonalize by a complex gauge
transformation. The problem is that if $\Phi$ commutes with its
adjoint in one frame, then it will generally not commute with its
adjoint in another frame that is related by a complexified gauge
transformation. But the frame depends on the choice of hermitian
metric, which must be solved for.

Fortunately, the existence of a solution to the non-abelian part of
the $D$-terms can still be phrased in algebro-geometric terms,
through a Higgs bundle analogue of the Uhlenbeck-Yau theorem. Let us
make some definitions. A subbundle $F \subset E$ is said to be a
Higgs subbundle if $\Phi(F) \subset F \otimes K$. A Higgs bundle is
said to be $J$-stable if
\be \mu(F) < \mu(E) \ee
for every Higgs subbundle, where the slope is defined as usual, $\mu
= J$-degree/rank. A Higgs bundle is semi-stable if $\mu(F) \leq
\mu(E)$ for every Higgs subbundle. Finally, a Higgs bundle is
poly-stable if it is a direct sum of stable Higgs bundles with the
same slope. Then the general principle is that the algebro-geometric
criterion of poly-stability should be equivalent to the differential
geometric criterion of the existence and uniqueness of the
hermitian-Einstein metric. (For abelian bundles, this requires
adding the explicit Fayet-Iliopoulos term $\zeta$ to the equation).

The condition of stability as a bundle is clearly stronger than the
condition of stability as a Higgs bundle. A Higgs bundle is stable
if the underlying bundle is. But a bundle which is blatantly
unstable can still be stable as a Higgs bundle. As a well-known
example, consider a Riemann surface $\Sigma_g$ with $g\geq 2$ and
choose a square root $K^{1/2}$ of the canonical bundle. Then take
\be E\ =\ K^{-1/2} \oplus K^{1/2}, \qquad \Phi\ =\ \left(
  \begin{array}{cc}
    0 & 1 \\
    0 & 0 \\
  \end{array}
\right) \ee
Then $E$ is unstable as a bundle, but stable as a Higgs bundle. The
slope of $E$ vanishes, and the slope of the Higgs sub-bundle
$K^{-1/2}$ is negative. The sub-bundle $K^{1/2}$, which destabilizes
$E$, is not preserved by the Higgs field.

Higgs bundles with generic spectral covers are stable. If the
spectral cover is smooth and irreducible, then the Higgs bundle
does not have any Higgs sub-bundles. In the next subsection we
would like to discuss stability when the spectral cover is not
smooth and irreducible.

Now we would like to compare this with the ALE fibration picture.
The cylinder mapping is a construction in algebraic geometry, so
the $F$-term data in the Higgs bundle description, the spectral
cover description and in the ALE fibration can be mapped exactly.
On the other hand, the hermitian-Einstein metric $h$ will not be
diagonal and the gauge field $A \sim h^{-1}\del h$ will be
non-abelian. This means that the $W$-bosons (i.e. the off-diagonal
components of the gauge field) will be condensed. On the other
hand, in the `closed string' ALE fibration picture the $W$-bosons
are extended solitons and do not have an off-shell description.
This means that a true ten/twelve dimensional solution does not
exist on the type II side, except perhaps in some fuzzy sense, as
the light $W$-bosons are not properly incorporated in the effective
action when the elliptic Calabi-Yau four-fold develops
singularities and cannot be condensed in this description. In the
Higgs bundle description the non-abelian degrees of freedom are
included, and this is why the Higgs bundle `resolves' the
singularities of the Calabi-Yau fourfold and provides a smooth
weakly coupled description.

This is a general phenomenon in heterotic/type II duality. The very
same phenomenon, in six dimensions instead of in eight dimensions,
was previously encountered very explicitly in
\cite{Wijnholt:2001us}. There we studied gravitationally dressed
versions of 't Hooft-Polyakov monopole solutions in type IIa on
$K3$. Such monopoles satisfy Bogomol'nyi equations, a close cousin
of Hitchin's equations. As described in \cite{Wijnholt:2001us}, the
abelian part of this solution (a Dirac monopole) is singular but can
be lifted to ten dimensions, where one encounters a $K3$ with $A_1$
singularity. To `resolve' the singularities, we incorporated
non-abelian gauge fields in the effective action. Then the
non-abelian part of the resulting solution smoothed out the
singularities of the abelian solution through exponentially small
corrections. But these corrections came from condensation of
extended solitons in $10d$. Such degrees of freedom cannot be
described by a local Lagrangian in $10d$, and thus the full
non-abelian solution could not be lifted. We encounter the same
problem here. One can try to integrate out the $W$-bosons and write
an abelian solution, which is singular at the branch locus of the
spectral cover. To smooth out the singularities however, we need to
include non-perturbative gauge fields, as in \cite{Wijnholt:2001us}.

As we noted in the introduction, this means that global models will
have difficulty capturing some non-abelian aspects of the local
model. In particular, in order to write down physical wave functions
and compute the K\"ahler potential in an approximation we can trust,
we need the hermitian-Einstein metric, which is not an object in
algebraic geometry and exists only in the Higgs bundle picture.

The question then arises if there isn't another way to deal with
the $D$-terms if we were working in the ALE fibration picture. Here
we can go back to the analogue of Uhlenbeck-Yau for Higgs bundles.
The criterion of existence and uniqueness of the hermitian-Einstein
metric can be phrased in terms of slope-stability. This is an
algebro-geometric concept which we could try to compare in the
Higgs bundle, spectral cover, and Calabi-Yau four-fold pictures. As
we discuss in more detail in the next section, slope-stability can
be defined in terms of Fayet-Iliopoulos parameters. According to
\cite{Donagi:2008kj,Donagi:2010pd}, the Fayet-Iliopoulos parameters
in $F$-theory are given by $\zeta_X \simeq m_{10}^4\int G \wedge J
\wedge \omega_X$. The difficult parts are (1) to properly define
all possible configurations of $C_3$ on a singular Calabi-Yau in
mathematical way, including the non-obvious configurations
considered in this paper, and (2) to define an analogue of the
notion of a sub and a quotient. This requires us to generalize the
notion of $\Ext^0$ for sheaves to ALE fibrations. These notions are
currently not available, but it is clear that some analogue should
exist at least in the context of ALE or del Pezzo fibrations,
because the map between spectral covers and ALE fibrations is an
algebraic one, so we can in principle define them by mapping to the
spectral cover side.

At any rate, given the relation with Higgs bundles it appears
inevitable that the primitiveness condition $J \wedge G=0$ must be
replaced by some notion of slope-stability for Calabi-Yau
four-folds with $G$-flux. Although in general we cannot write the
physical wave functions in the Calabi-Yau four-fold or spectral
cover pictures, if phrased in such terms, the essential information
of existence and uniqueness of the hermitian-Einstein metric can be
preserved. This is an important qualitative change, because
primitiveness is a closed condition, whereas stability is an open
condition and leads to a chamber structure in the K\"ahler moduli
space.

We emphasize again that this situation is not unique to $F$-theory.
For example in the context of $M$-theory on $G_2$-manifolds, the
$G_2$-metric is singular near the three-cycle where the gauge
theory is localized and is therefore not the correct metric for
physics purposes. To understand the physics near such three-cycles,
we need some way to `resolve' the singularities and obtain a smooth
weakly coupled description. This was achieved only recently in
\cite{Pantev:2009de}, by replacing the singular $G_2$ metric by the
harmonic metric on a Higgs bundle. The harmonic metric is smooth
and includes non-abelian corrections, but again only exists in the
Higgs bundle picture. Thus even when new techniques for
constructing compact $G_2$-holonomy manifolds become available, in
the regime of interest the $G_2$-metric can't be trusted and we
still have to go back to the local model in order to study the
$D$-terms (using the harmonic metric).

\newsubsection{Stability for degenerate cases}

\subseclabel{SimpsonStability}

In the previous subsection we rephrased the existence and uniqueness
of the hermitian-Einstein metric in terms of slope-stability of the
Higgs bundle. Because Higgs bundles are usually constructed by
writing down spectral data, it would be more convenient to have a
stability criterion for the spectral sheaf. However we have seen
that the spectral data for a smooth Higgs bundle can easily have
singular behaviour, for example the spectral cover can be reducible
or non-reduced. Thus we need a criterion that behaves well under
degenerations, and remains equivalent to existence and uniqueness of
the hermitian-Einstein metric in the Higgs bundle picture even in
such degenerate cases.

The theory of stable sheaves is generally credited to Gieseker,
Maruyama, and Simpson \cite{SimpsonPi1Moduli}, and is based on the
Hilbert polynomial. The Hilbert polynomial is defined purely
algebraically and is constant in flat families, even if some members
of the family are degenerate. Physically speaking this implies for
instance that the net number of generations cannot jump.

Thus instead of a K\"ahler class $J$, we consider an ample line
bundle $\cO(1)$ whose first Chern class is proportional to $J$. Then
we consider the associated Hilbert polynomial
\be P_{\rm Hilb}({\cal L}, m) \ = \ \chi({\cal L} \otimes \cO(m))
\ee
where of course $\chi(F) = \sum (-1)^i \Ext^i(\cO_X, F)$. We define
the coefficients
\be P_{\rm Hilb}({\cal L}, m) \ = \ \sum_{k=0}^d p_k({\cal L})
{m^k\over k!} \ee
Using Riemann-Roch, they can be expressed in terms of Chern classes.
The degree of $P_{\rm Hilb}({\cal L}, m)$ is the dimension of the
support of ${\cal L}$, and the coefficient $p_i$ of the leading term
is called the rank. In our case, we will be interested in sheaves
that are supported in dimension two on a three-fold, so $p_3=0$ and
$d=3$. Then, the slope is defined as
\be \mu({\cal L}) \ = \  {  p_1({\cal L}) \over  p_2({\cal L}) } \ee
and slope-stability is defined in the usual way. Note that this
makes sense for arbitrary coherent sheaves on a projective variety,
in particular reducible or non-reduced cases. (There is also the
notion of Gieseker stability, which uses the normalized Hilbert
polynomial $p({\cal L},m) = P({\cal L},m)/{\rm rank}$ instead of the
slope, but we do not know how to justify this in the context of
$F$-theory).

To apply this to our case, we let $\bar X$ be the projective closure
of $X$. We may pick an ample $\cO(1)_{\bar X}$ which restricts to
$\cO(1)$ on $X$. Then, provided the spectral cover does not
intersect infinity, slope-stability for the spectral sheaf is the
same as slope-stability for the Higgs bundle. To see this, K\"ahler
classes on $\bar X$ are of the form
\be  \pi^*J_{B} + t J_0 \ee
where $J_B$ is a class on the base, $J_0$ is the Poincar\'e dual of
the zero section, and $t$ is a real number. Restricting to $X$, the
class $J_0$ trivializes, and we are left with $\pi^*J_B$. Then
stability of ${\cal L}$ with respect to $\pi^*J_B$ is the same as
stability of $E = p_{C*}{\cal L}$ with respect to $J_B$ (or any
multiple of it), because $H^i(C, {\cal L}\otimes (p_C^*L)^m) \cong
H^i(S,E \otimes L^{m})$. But $\pi^*J_B$ is not a K\"ahler class on
$\bar X$. To fix this, we consider a small perturbation by $\epsilon
J_0$. Since stability is an open condition, a sufficiently small
perturbation preserves stability. Then by rescaling $J_B + \epsilon
J_0$ and relabelling $J_B$, we see that stability of ${\cal L}$
agrees with stability for the Higgs bundle. Slope-stability is also
usually preserved under the Fourier-Mukai transform, for instance in
the context of heterotic spectral covers \cite{Andreas:2003zb}.

We can also adapt these statements when there is a parabolic
structure. One may define a slope for parabolic sheaves, and use
this to define stability for the spectral sheaf.

Thus stability for sheaves gives a practical way to see if the
$D$-terms are satisfied. In particular this gives a simple way to
derive a statement in the previous subsection: generic Higgs
bundles, for which the spectral sheaf is actually an honest line
bundle, are stable. This follows simply because any line bundle is
stable.

From Riemann-Roch we get
\be
 p_2\ =\  {\rm ch}_1({\cal L})J^2, \qquad
  p_1\ =\  {\rm ch}_2({\cal L})J + {\rm ch}_0({\cal L}){c_2(TX)\over 12} J
\ee
For the special case of a bundle $L$ on a divisor $D$, ${\cal L} =
i_*L$, we have ${\rm ch}_0 = 0$, ${\rm ch}_1 = {\rm rank}(L)\, D$,
and ${\rm ch}_2 = i_{D*}c_1(\hat L)$, leading to
\be
  p_2\ =\ {\rm rank}(L) \int_D J \wedge J, \qquad
  p_1\ = \
  \int_D J \wedge c_1(\hat L)
\ee
where $\hat L = L \otimes K_D^{-1/2}$, and so the expression for the
slope reduces to the usual one.

The Chern characters for several configurations of interest were
discussed in section \ref{Index}, and can easily be used to write
down the slope. For instance for the reducible case, where ${\cal
L}$ is given by an extension
\be 0 \ \to \ i_{2*}L_2(-\Sigma)\ \to \ {\cal L} \ \to \ i_{1*}L_1\
\to \ 0 \ee
the Chern character of ${\cal L}$ is given in equation
(\ref{ReducibleChern}), and therefore the slope is given by
\be \mu({\cal L})\ =\ {\deg(\hat L_1) + \deg(\hat L_2) - \int_\Sigma
J \over {\rm vol}(D_1) + {\rm vol}(D_2)} \ee
In this case, $i_*L_2(-\Sigma)$ is clearly a potential destabilizing
subsheaf, whereas $i_*L_1$ is not a subsheaf.

In the effective theory, the slopes are closely related to field
dependent Fayet-Iliopoulos terms.%
\footnote{We effectively use the old supergravity arguments, which
assume only $N=1$ supersymmetry. More recent work often uses a
`central charge' function. This assumes a broken underlying $N=2$
supersymmetry and is therefore less
general.} %
Let us consider for example a configuration of $7$-branes ${\cal L}$
in type IIb. We are interested in the reducible case, i.e. we have
${\cal L} = \oplus_n {\cal L}_n$ where the ${\cal L}_n$ are
irreducible. The abelian generators of the low energy gauge group
are given by $\Ext^0({\cal L}_n, {\cal L}_n)$ and the non-abelian
generators are given by $\Ext^0({\cal L}_m,{\cal L}_n)$ with $n \not
= m$. Suppose the low energy gauge group is $G$, and let $\xi:G \to
U(1)$ be a character. At the level of the Lie algebra, it
corresponds to a linear combination $\sum_n\xi_n \omega_n$ where
$\omega_n$ is the generator of $\Ext^0({\cal L}_n,{\cal L}_n)$. To
each such $\xi$ we associate a twisted version of the Chern
character. Intuitively we think of this as the Chern character of a
rank one sheaf ${\cal L}_\xi$ corresponding to the $U(1)$ gauge
symmetry, although ${\cal L}_\xi$ may strictly not exist:
\be {\bf ch}( {\cal L}_\xi) \ = \ \sum_n \xi_n\, {\bf ch}( {\cal
L}_n)\ee
Then for each $\xi$ we get a shift symmetry on the K\"ahler moduli
space
\be \delta_{\lambda_\xi}\, {\rm Im}(T_D)\ =\ [\omega^{(2)}({\cal
L}_\xi, \lambda_\xi) ]_2 \cdot D \ \simeq\ \lambda_\xi\,{\rm
ch}_2({\cal L}_\xi)\cdot D \ee
where $\omega^{(2)}({\cal L}_\xi, \lambda_\xi)$ is obtained by
descent:
\be {\bf ch}( {\cal L}_\xi){\bf A}^{1/2}(X)\  =\ d\,
\omega^{(1)}({\cal L}_\xi), \qquad \delta_{\lambda_\xi}
\omega^{(1)}\ = \ d\,\omega^{(2)}({\cal L}_\xi,\lambda_\xi) \ee
In general such a shift symmetry is deduced from the Chern-Simons
couplings of the gauge fields to anti-symmetric tensor fields. In
IIb this follows from the Chern-Simons coupling of ${\cal L}$ to
$C^{(4)}_{RR}$. Although the expression was derived for smooth
configurations, in this form it applies equally well to general
coherent sheaves, like the reducible or non-reduced configurations
considered in this paper, or even a complex of such. The reason is
that we can resolve each ${\cal L}_n$ as a sequence of vector
bundles. Since the Chern character is additive, we apply descent to
the individual pieces, and then we add them back together with
appropriate signs. In the heterotic setting, we get essentially the
same story by considering the transformation law for $B^{(2)}_{NS}$
and $\tilde B^{(6)}_{NS}$, and in IIa we would consider the
transformation law for $C^{(3)}_{RR}$.

Apart from an isometry, to define a moment map we further need a
K\"ahler form on the moduli space. This is also determined by the
string compactification. In the large volume limit, it can be
determined by a Kaluza-Klein reduction, and in type IIb for example
is given by the second derivative of the following K\"ahler
potential:
\be {\cal K}\ =\ -2M_{Pl}^2 \log {\cal V} 
\ee
where ${\cal V}$ is the volume as a function of the K\"ahler moduli.
With this potential, the Fayet-Iliopoulos parameter (or moment map
for the Killing vector field of the shift symmetry) is precisely
given by the slope $\mu({\cal L}_\xi)$, up to an over-all factor
which is moduli dependent but independent of the details of the
brane.

We would like to reexamine the mathematical notion of slope
stability in light of this relation between the slope and the
Fayet-Iliopoulos parameter in the effective Lagrangian. It helps to
generalize slightly and consider an abstract brane $E$, which can
be a bundle, a coherent sheaf, a Lagrangian submanifold or a
boundary state depending on the context. In this article we have
argued it must be even further extended to ALE fibrations. Then we
have the following well-known and universal phenomenon in string
compactification.

We adjust the K\"ahler moduli until $E$ becomes marginally stable to
decay into two subobjects, $E'$ and $E''$. Then one finds that at
the locus of marginal stability, the effective theory is described
by a version of the Fayet model
\cite{Sharpe:1998zu,Kachru:1999vj,Douglas:2000ah,ThomasMirrorStability}.
That is, first of all we get an extra $U(1)$ gauge symmetry
$U(1)_X$, equivalently an extra generator
\be \Lambda_X\ \in\ \Ext^0(E,E) \ee
This is practically the definition of marginal stability. At the
wall of marginal stability, $E$ becomes semi-stable, and the
solution of the $D$-terms yields the unique reducible object with
the same graded sum, $E \sim E' \oplus E''$. Then $\Ext^0(E,E)$ is
at least two-dimensional, with $\Lambda' \in \Ext^0(E',E')$ and
$\Lambda'' \in \Ext^0(E'',E'')$, and we identify $\Lambda_X =
\Lambda' - \Lambda''$. Secondly, we get an extra generator $X \in
\Ext^1(E,E)$, i.e. a chiral field $X$ in $\Ext^1(E',E'')$ or
$\Ext^1(E'',E')$. From the Yoneda pairing $\Ext^0 \times \Ext^1 \to
\Ext^1$, we see that the chiral field is charged under $U(1)_X$,
i.e. we have
\be \delta X\ =\ \Lambda' X - X \Lambda'' \ee
When $X$ gets a VEV, we see that $ X \sim X + \Lambda_X$, so $X$
becomes exact and $\Lambda_X= \Lambda' -\Lambda''$ is no longer
closed, and both are removed from the massless spectrum. Using the
Hermitian metric to separate complexified gauge transformations in
actual gauge transformations and $D$-terms, this is equivalent to
saying that the $U(1)$ is Higgsed, and we have a $D$-term potential
of the form
\be V_D \ =\ \half (\zeta_X - q_X |X|^2)^2 \ee
which is a version of the Fayet model.

Now let us connect this with the notion of slope stability. We
regard $F$ as a non-trivial extension of $E'\oplus E''$. Then the
relevant $U(1)$ symmetry is $\rho_\xi = \Lambda' - \Lambda''$, so we
have
\be \zeta_X\ =\ \mu(E') - \mu(E'') \ee
From the $D$-term potential of the Fayet model, when $\zeta_X
> 0$ we find that $X$ gets a VEV and we form a bound state. When
$\zeta_X = 0$ there is a supersymmetric vacuum with $\vev{X}=0$ and
massless $U(1)_X$. And when $\zeta_X < 0$, supersymmetry is broken
by $D$-terms. Now it is not hard to prove that if $F$ is given by an
extension
\be\label{FExtDef} 0 \to E'' \to F \to E' \to 0 \ee
then we have either $\mu(E'') < \mu(F) < \mu(E')$ or $\mu(E'') >
\mu(F) > \mu(E')$. Assuming there are no other light fields in the
$D$-term potential, it follows that $F$ is stable for $\zeta_X >
0$, $E' \oplus E''$ is poly-stable for $\zeta_X = 0$, and the
system is unstable for $\zeta_X < 0$. This seems to agree nicely
with our discussion of slope stability.

However, there is an important subtlety in the above discussion,
which seems to be ignored in the literature. What we really want to
consider is infinitesimal deformations, i.e. deformations over the
dual numbers $D = {\bf C}[\varepsilon]/\varepsilon^2$ (see eg.
\cite{HartshorneDef}). Physically the reason for this is that the
Fayet model is only an effective description for the linearized
deformations. We could certainly also consider finite deformations,
but stability is a highly non-linear condition and the Fayet model
could hardly be expected to capture this. In fact for intersecting
brane configurations, there are always at least two natural and
inequivalent quotient branes, given by restricting to either of the
two intersecting components. It is not hard to see that they give
inequivalent restrictions on the slope. Therefore for a finite
deformation we get at least two inequivalent `decay modes' for
which we have to test stability, whereas the Fayet model sees only
one. Ignoring the second `decay mode' quickly leads to
contradictions with Murayama's boundedness result. But it seems
natural to conjecture that in generic enough situations, testing
against these two decay modes should be sufficient to ensure
stability for a finite deformation. This will be the de facto
assumption in some of the examples in part II.

\newsubsection{Numerical approach with balanced metrics}

\subseclabel{NumMetric}

As usual in supersymmetric string compactification, the zero modes
and superpotential can be determined up to field redefinitions by
methods of algebraic geometry. As we discussed in detail, even the
existence of a solution of the $D$-term equations can be
characterized in algebro-geometric terms. However, for certain
questions existence does not suffice, and we need to have an
explicit knowledge of the physical wave-functions. This is necessary
to understand detailed flavour structure originating in the K\"ahler
potential, or more accurate predictions for dimension six proton
decay \cite{Donagi:2008kj}. For this, we need to map wave-functions
derived in the holomorphic frame back to a unitary frame, i.e. we
need to find $H$. Actually all physical quantities depend only on
$H$ up to ordinary $SU(n)$ gauge transformations, and they can be
expressed using the hermitian-Einstein metric $h$. So we need to
explicitly solve for $h$.

Thus the question arises how we get a handle on this. As we saw
above, the hermitian-Einstein metric satisfies a non-linear elliptic
PDE which is virtually impossible to solve explicitly.

In the analogous problem of finding solutions to the hermitian
Yang-Mills equations on a complex vector bundle, the situation has
improved in recent years by the development of numerical
approximation schemes for the Hermitian Yang-Mills metric
\cite{DonaldsonBalanced,WangCanonical,Douglas:2006hz,Anderson:2010ke}.
This is based on many standard ideas in geometric invariant theory.
We will briefly review some of the ingredients below and then
conjecture a natural analogue for approximating the
hermitian-Einstein metric on Higgs bundles over K\"ahler manifolds.
The latter can then be applied to brane configurations in type II
settings, as long as the field theory approximation applies. This
includes type IIb and $F$-theory compactifications, in the limit
that the angles between intersecting branes are small. A modified
version should also apply to type IIa and $M$-theory
compactifications, where one needs to approximate the harmonic
metric \cite{Pantev:2009de}, and type I' compactifications, where one
studies a generalized version of monopole equations
\cite{PW-p}.\footnote{$D$-term structure in Higgs bundles has been
studied recently eg. in \cite{Hayashi:2009bt}, however no systematic
approximation scheme was specified there.}

Let us consider a Calabi-Yau $d$-fold $Z$ with a holomorphic bundle
$V$ of rank $r$ and $c_1(V) = 0$. We are interested in solutions of
\be g^{i\bar j}F_{i\bar j} \ = \ 0 \ee
which we interpret as an equation for the hermitian metric $h$ on
$V$. The solution is called the hermitian Yang-Mills metric or the
hermitian-Einstein metric. We will use the former terminology in
order to distinguish between the hermitian-Einstein equation for a
Higgs bundle, which has an extra term proportional to
$[\Phi,\Phi^\dagger]$.

The hermitian Yang-Mills metric on a bundle $V$ of rank $r$ may be
approximated by a sequence of balanced metrics. The idea is as
follows. We consider an ample line bundle $L$, in fact we will take
$L$ to be the ample line bundle for which $c_1(L)$ is the K\"ahler
form $J$. For large enough $m$,  $H^0(V\otimes L^m)$ is generated by
sections $s_u$, $u = 1, \ldots ,N$, and the higher cohomologies
vanish. These sections then define an embedding map
\be i:\, Z\ \to\ {\rm Gr}(r,N) \ee
We have the tautological rank $r$ bundle $U_r$ over $ {\rm
Gr}(r,N)$, whose fiber over an $r$-plane in ${\bf C}^N$ is given by
the $r$-plane itself, and we have $V\otimes L^m = i^*U_r^\vee$. Now
let us pick an $N\times N$ matrix $M^{u\bar v}$, defining a
Fubini-Study metric for $U_r$. For each such matrix, we get a
hermitian metric $h_M$ on $V\otimes L^m$ by pulling back:
\be\label{FSmap} (h^{-1}_M)^{a\bar b}\ =\ s_u^a \,M^{u\bar v}
(s^\dagger)_{\bar v}^{\bar b} \ee
By subtracting the trace, this yields a Hermitian metric on $V$. The
space of inequivalent metrics we get this way, or alternatively the
space of inequivalent embeddings into ${\rm Gr}(r,N)$, is
parametrized by $Sl(N,{\bf C})/SU(N)$. In particular, these metrics
are algebraic, can be written down explicitly as above once we have
a basis of holomorphic sections, and depend only on a finite number
of parameters in the matrix $M$, whereas a general hermitian metric
on $E$ depends on infinitely many parameters and is not algebraic.
Thus the idea is to find the best approximation to the hermitian
Yang-Mills metric within this finite dimensional space of algebraic
metrics, and then increase $m$ to make the error as small as one
wishes.

Thus our task is to produce the best metric of the form $h_M$. For
this we proceed as follows. Given an arbitrary hermitian metric $h$
on $V\otimes L^m$ (not necessarily of the form $h_M$), we have the
natural $L^2$ inner product on the space of sections, which
restricts to an inner product $M$ on the space of global sections
$H^0(Z,V)$ given as
\be\label{Hilbmap} (M^h)^{-1}_{u\bar v} \ = \ \int_Z \vev{s_u,
s_v}_h\, d{\rm vol} \ee
where $d{\rm vol} = J^d/d!$ is the volume form defined by the
K\"ahler metric $g$ on $Z$. Now let us take $\{s_i\}$ to be a basis
of $H^0(Z,V)$ which is ortho-normal with respect to $M^h$. Assuming
$V\otimes L^m$ is generated by global sections, we can define the
Bergman kernel as
\be B_h \ = \ \sum_{i=1}^N s_i \otimes s_i^{\dagger_{M^h}} \ \in \
C^{\infty}(Z,{\rm End}(V\otimes L^m)) \ee
In other words, it corresponds to orthogonal projection on the zero
mode sector. The kernel does not depend on the specific choice of
ortho-normal basis. The trace of the kernel is given by
\be  {\rm Tr}(B_h)\ = \ N\ =\   \chi_{Hilb}(V,m) \ee
where $\chi_{Hilb}(V,m)$ is the Hilbert polynomial with respect to
$L$:
\be \chi_{Hilb}(V,m)\ = \ r\cdot {\rm vol}(Z)\, m^d + \left({\rm
deg}(V) + {r\over 2} {\rm deg}(TZ) \right) m^{d-1} + \ldots \ee
This follows because as we said before, the line bundle $L$ is
positive and so the higher cohomologies of $V \otimes L^m$ all
vanish for large enough $m$. Furthermore, the kernel has the
following asymptotic expansion:
\be\label{BhExpansion} \left| B_h - m^d {\bf 1}_{r\times r} -\left(
\half R_g\,{\bf 1}_{r\times r}  +\sqrt{-1}\, g^{i\bar j}F_{i\bar
j}\right) m^{d-1} \right|\ \leq\  C \, m^{d-2} \ee
where $R_g$ is the scalar curvature for the metric $g$, $R_g \sim
-\sqrt{-1} g^{i\bar j}\del_i\del_{\bar j} \log \det (g)/2\pi$. Of
course for a Calabi-Yau metric (which can be found by similar
methods), we would have $R_g$ and $c_1(Z)$ vanishing. Keeping the
trace part around would not problematic, because hermitian metrics
on line bundles are relatively simple and we can easily correct for
them, but let us assume they vanish for simplicity. Therefore, we
see that if we can find a metric $h$ for which $B_h$ is constant, or
more precisely $B_h = \chi(V,m){\bf 1}_{r\times r}/(r\cdot {\rm
vol}(Z))$, then we have
\be \left| {\deg(V)\over r\cdot {\rm vol}(Z)}{\bf 1}_{r\times
r}\,m^{d-1} -\sqrt{-1}\left(g^{i\bar j}F_{i\bar j}\right) m^{d-1}
\right|\ \leq\  \tilde C \, m^{d-2} \ee
In other words, the error with this choice of metric $h$ scales as
$1/m$, and for large $m$ we approximate the Hermitian Yang-Mills
metric arbitrarily well. A metric for which the Bergman kernel is
constant is said to be balanced, at least this is one of several
equivalent definitions.

So, we need a metric $h_M$ which is balanced. To find this metric,
we can use an iteration procedure. We had the assignment
\be FS: M \to h_M \ee
in (\ref{FSmap}). In other words, if we think of $M$ as
parametrizing embeddings, we pull back the Fubini-Study metric on
$U_r^\vee$. Conversely, we saw in (\ref{Hilbmap}) that we had the
assignment
\be {\rm {\cal H}ilb}: h \to M^h \ee
Thus given a matrix $M$, we have an operator
\be T(M)\ =\ {\rm {\cal H}ilb} \circ FS(M) \ee
Concretely, we have the formula
\be T(M)_{u\bar v}^{-1}\ =\ {N\over {\rm vol}(Z)\, r} \int_Z
{s_{\bar v}^\dagger }  \, h_M\,s_u \ d{\rm vol} \ee
This produces a sequence $M_{i+1} = T(M_i)$, equivalently a sequence
in $Sl(N,{\bf C})/SU(N)$. The fixed point $M_{\infty} =
T(M_{\infty})$ yields the balanced metric, and if the balanced
metric exists (which happens if $V$ is stable), then the sequence
converges to it. In practice a few iterations yield a good
approximation.

Incidentally, there is a sense in which balanced metrics may be
regarded as quantized versions of hermitian-Einstein metrics, with
$\hbar = 1/m$ \cite{Lukic:2007nc,IuliuLazaroiu:2008pk}. It is
currently not completely clear to us what the significance of this
is in the context of phenomenological string compactifications (see
\cite{Douglas:2008pz,Douglas:2008es} for a possible interpretation
in a slightly different setting), but it would surely be interesting
if balanced metrics have some physical significance beyond serving
as approximations of hermitian-Einstein metrics. Also, the existence
of balanced metrics is equivalent to Gieseker stability, which uses
the full Hilbert polynomial and is stronger than the slope-stability
we have used. This suggests that a small modification of the
balanced metric yields a solution to the deformed hermitian
Yang-Mills equations studied by Leung \cite{LeungGStable}. This is
also closely related to the $\alpha'$ corrected version of the
(abelian) hermitian Yang-Mills metric in type II settings, studied
in \cite{Marino:1999af,Enger:2003ue}. In the context of the
heterotic string it seems to be closely related to a $g_s$-corrected
version of the slope \cite{Dine:1987xk,Blumenhagen:2005ga}. One can
presumably investigate this by considering subleading terms in the
expansion of the Bergman kernel (\ref{BhExpansion}).

We need an extension of this story for Higgs bundles. This does not
seem to have been stated in the literature, but the following
proposal is closely related to \cite{LWangVortex,KellerVortex}. We
will assume that the Higgs bundle $(E,\Phi)$ is defined over a
K\"ahler manifold (as in $F$-theory or IIb, but not in $M$-theory,
IIa or type I') and does not have poles. Further adjustments may
have to be made when the Higgs field is meromorphic.

Our proposal is the following modification. We still want to use the
metrics above to approximate the Hermitian-Einstein metric, or at
least a closely related set of metrics parametrized by the same
finite dimensional space, so again we pick a positive line bundle
$L$ (with $c_1(L) = J$) and consider the space of sections
$H^0(S,E\otimes L^m)$ in order to get an embedding into $Gr(r,N)$,
with $r = {\rm rank}(E)$ and $N = h^0(S,E\otimes L^m)$. But we will
have to modify the balance condition in a $\Phi$-dependent way. The
idea will be to change the balance condition by terms of order
$1/m$. Note the balanced metric itself may not even exist, as Higgs
bundles which are stable can be highly unstable as ordinary bundles.
Then the curvature $g^{i\bar j}F_{i\bar j}$ is modified at order
$1/m$, so this leads only to an order $m^{d-2}$ correction to
(\ref{BhExpansion}) which can be absorbed in $\tilde C$. Similarly,
the Bergman kernel is modified at order $1/m$. Inspecting
(\ref{BhExpansion}), we see that we do not want a metric $h$ for
which $B_h$ is constant, but instead we want a metric $h'$ for which
\be\label{PhiBalanced} B_{h'}\ =\ {\chi(E,m)\over r\,{\rm vol}(S)}
{\bf 1}_{r\times r} -\sqrt{-1}\, m^{d-1}\, [\Phi^{\dagger_{h'}},
\Phi] \ee
In fact, for our purposes this only needs to hold up to terms of
order $m^{d-2}$.

The above observations tell us how to modify the balance condition
by terms of order $1/m$. We modify the inner product (\ref{Hilbmap})
in the following way:
\be\label{PhiBalMetric} \vev{s_u,s_v}_{FS(M)}\ =\ \vev{({\bf 1} +
{\bf \epsilon})\,s_u,s_v}_{h_M} \ee
where $\epsilon$ is of order $1/m$, and is itself $h$-dependent. We
need to ensure that (\ref{PhiBalMetric}) actually defines a metric,
which seems to be fine for large $m$. This will have to be
reexamined when we allow for poles of the Higgs field. Using the new
definition of the map $FS$, we can propose a new $T$-operator as $T
= {\rm {\cal H}ilb} \circ FS$. Concretely it is given by
\be T(M)^{-1} \ = \ {N\over {\rm vol}(S)\, r} \int_S s^\dagger
(sMs^\dagger)^{-1} (1 +\epsilon) s \, d{\rm vol}\ee
where $\epsilon$ itself will be defined using $h_M =
(sMs^\dagger)^{-1}$. At a fixed point $T(M_{\infty}) = M_{\infty}$
it is convenient to make a change of basis so that $M_{\infty} =
I_{N\times N}$. If $s$ is the corresponding embedding, then $s$ is
an ortho-normal basis for $FS(M_{\infty})$, and thus can be used to
write down the Bergman kernel for $FS(M_{\infty})$. Now $s$ is not
an ortho-normal basis for $h_{M_\infty} = (ss^\dagger)^{-1}$, but we
can still consider the projection operator
\be P_{h_M}\ =\ s s^{\dagger_{h_M}}\ =\ s s^\dagger (s s^\dagger
)^{-1}\ =\ {\bf 1}_{r\times r} \ee
and the metric $FS(M)$ differs from $h_M$ by $FS(M) = (sM
s^\dagger)^{-1}(1+\epsilon)$. Thus given a solution of the fixed
point equation, we find that the Bergman kernel for $FS(M_\infty)$
satisfies
\be B_{FS(M_\infty)} \ = \ {N\over r\,{\rm vol}(S)}\, s\,
s^{\dagger_{h_{M_\infty}}}(1 + \epsilon)\ = \ {N\over r\,{\rm
vol}(S)}\,({\bf 1}_{r\times r} + \epsilon)  \ee
We see that if we take
\be  \epsilon \ =\   - \sqrt{-1}\,{ r\, {\rm vol}(S)\ m^{d-1}\over
N}\,[\Phi^{\dagger_{h_{M_\infty}}}, \Phi] \ee
then the Bergman kernel for $FS(M_\infty)$ gives the desired
expression (\ref{PhiBalanced}) with $h' = FS(M_\infty)$ up to terms
of order $m^{d-2}$. Although we derived this statement in a basis
such that $M_{\infty} = I_{N\times N}$, it is independent of this
choice. Let us call such metrics $\Phi$-balanced.

Using the new $T$-operator, we manufacture a sequence by applying
the $T$-operator, $M_{i+1}=T(M_i)$. The main gap is that we have not
given an argument that a unique fixed point exists and that the
sequence converges to it. By analogy with conventional balanced
metrics, we may conjecture that a unique fixed point exists if the
Higgs bundle is stable. The $\Phi$-balanced metric $FS(M_\infty)$
then gives an approximation to the hermitian-Einstein metric on the
Higgs bundle $E \otimes L^m$, converging to it in the limit $m \to
\infty$. By subtracting the trace, we get an approximation to the
hermitian Einstein metric on $E$ itself.

Eventually one should also take into account that the Higgs bundles
appearing in $F$-theory are meromorphic. There is a moment map
formulation for the D-terms of a parabolic Higgs bundle, so in
principle the story above could be adjusted to this case.
Alternatively, one could investigate Higgs bundles over surfaces
where $K_S$ is positive, or work with $K(D)$ valued Higgs fields,
which will presumably yield similar qualitative behaviour.

\newpage

\bigskip

\noindent {\it Acknowledgements:} We would like to thank
L.~Anderson for initial collaboration, T.~Pantev for discussions,
and D.~Joyce for correspondence related to this project. MW would
like to thank C.~Vafa and the Heisenberg program of the German
Science Foundation for generous financial support during this
project. RD acknowledges partial support by NSF grants 0908487 and
0636606. MW would also like the Taiwan string theory workshop and
the University of Bonn for the opportunity to present some of these
results.

\end{document}